\documentclass[lettersize,journal]{IEEEtran}
\usepackage{stfloats}
\usepackage{url}
\usepackage{verbatim}
\usepackage{array}
\usepackage{cite}
\usepackage{amsmath,amssymb,amsfonts}
\usepackage{gensymb}
\usepackage{amsthm}

\usepackage{booktabs}
\usepackage{array}
\usepackage{adjustbox}
\usepackage{tabularray}
\usepackage{diagbox}
\usepackage{colortbl}
\usepackage{bbding}
\usepackage{tabularx}

\usepackage{graphicx,subfigure}
\usepackage{textcomp}
\usepackage{xcolor}
\usepackage{flushend}
\usepackage{makecell}
\usepackage{caption}
\usepackage{float}

\usepackage[linesnumbered,ruled,vlined]{algorithm2e}

\usepackage{comment}

\begin{document}

\title{Joint Task Offloading and Resource Allocation in Low-Altitude MEC via Graph Attention Diffusion}

\author{Yifan Xue, Ruihuai~Liang,~\IEEEmembership{Student Member,~IEEE,} 
Bo~Yang,~\IEEEmembership{Member,~IEEE,} Xuelin Cao,~\IEEEmembership{Member,~IEEE,} \\ Zhiwen Yu,~\IEEEmembership{Senior Member,~IEEE,}   M\'erouane Debbah,~\IEEEmembership{Fellow,~IEEE}, 
and Chau~Yuen,~\IEEEmembership{Fellow,~IEEE}  
\thanks{

Y. Xue, R. Liang, and B. Yang are with the School of Computer Science, Northwestern Polytechnical University, Xi'an, Shaanxi, 710129, China (email: liangruihuai$\_$npu@mail.nwpu.edu.cn, yang$\_$bo@nwpu.edu.cn). 

Z. Yu is with the School of Computer Science, Northwestern Polytechnical University, Xi'an, Shaanxi, 710129, China, and Harbin Engineering University, Harbin, Heilongjiang, 150001, China (email: zhiwenyu@nwpu.edu.cn).

X. Cao is with the School of Cyber Engineering, Xidian University, Xi'an, Shaanxi, 710071, China (email: caoxuelin@xidian.edu.cn). 

M. Debbah is with the Center for 6G Technology, Khalifa University of Science and Technology, P O Box 127788, Abu Dhabi, United Arab Emirates (email: merouane.debbah@ku.ac.ae). 



C. Yuen is with the School of Electrical and Electronics Engineering, Nanyang Technological University, Singapore (email: chau.yuen@ntu.edu.sg). 

 }
\vspace{-0.7cm}
}

\markboth{Journal of \LaTeX\ Class Files,~Vol.~XX, No.~XX, December~2024}%
{Shell \MakeLowercase{\textit{et al.}}: A Sample Article Using IEEEtran.cls for IEEE Journals}


\maketitle

\begin{abstract}
With the rapid development of the low-altitude economy, air-ground integrated multi-access edge computing (MEC) systems are facing increasing demands for real-time and intelligent task scheduling. In such systems, task offloading and resource allocation encounter multiple challenges, including node heterogeneity, unstable communication links, and dynamic task variations. To address these issues, this paper constructs a three-layer heterogeneous MEC system architecture for low-altitude economic networks, encompassing aerial and ground users as well as edge servers. The system is systematically modeled from the perspectives of communication channels, computational costs, and constraint conditions, and the joint optimization problem of offloading decisions and resource allocation is uniformly abstracted into a graph-structured modeling task. On this basis, we propose a graph attention diffusion-based solution generator (GADSG). This method integrates the contextual awareness of graph attention networks with the solution distribution learning capability of diffusion models, enabling joint modeling and optimization of discrete offloading variables and continuous resource allocation variables within a high-dimensional latent space. We construct multiple simulation datasets with varying scales and topologies. Extensive experiments demonstrate that the proposed GADSG model significantly outperforms existing baseline methods in terms of optimization performance, robustness, and generalization across task structures, showing strong potential for efficient task scheduling in dynamic and complex low-altitude economic network environments.
\end{abstract}


\section{Introduction}
\IEEEPARstart{I}{n} recent years, the rapid advancement of wireless communication and manufacturing technologies has significantly facilitated the widespread adoption of unmanned aerial vehicles (UAVs) and other aerial platforms. Coupled with the increasing congestion of terrestrial road infrastructures, these developments have catalyzed the emergence of the Low-Altitude Economy (LAE). LAE focuses on utilizing low-altitude airspace (below 3,000 meters above sea level) to support a wide range of societal activities, such as aerial traffic management, logistics and delivery services, and airborne surveillance, thereby giving rise to Low-Altitude Economy Networks (LAENet). These networks serve not only as carriers of critical communication tasks but also as infrastructures that integrate computation, storage, and sensing services, forming a fundamental backbone for the development of LAE \cite{zhao2025generativeaienabledwirelesscommunications}. The primary goal of LAE is to leverage aerial vehicles to enhance the efficiency of social operations and to stimulate economic vitality through optimized resource allocation. Advanced technologies such as UAVs, electric vertical take-off and landing (eVTOL) aircraft, and satellite communications play pivotal roles in this ecosystem. Through intelligent network collaboration, these technologies have enabled applications in various sectors, including logistics, agriculture, emergency response, and urban governance\cite{zheng2025uavswarmenabledcollaborativepostdisaster}. 

With the rapid development of the low-altitude economy and mobile applications, the number of network nodes has increased, task demands have become more complex\cite{10964011,10577218}, and their dynamics have intensified. However, user equipment (UE), such as UAVs, often lacks sufficient computational capacity and battery endurance, making it difficult to handle these computation-intensive and latency-sensitive tasks independently. To address these challenges, multi-access edge computing (MEC) has emerged as an extension of cloud computing to support data processing and communication within mobile networks \cite{9725258}. Unlike traditional approaches that transmit computation requests to centralized data centers, MEC deploys powerful computing resources at the network edge, thereby significantly enhancing both communication and computational capabilities. Through wireless connections, UEs can offload computation tasks to nearby edge servers, enabling efficient local processing and substantially reducing energy consumption \cite{10373079}.

Traditional MEC systems deployed in wireless cellular networks rely on terrestrial infrastructure, where ground UEs generate computation tasks and offload them via wireless connections to nearby base stations for processing. However, in remote areas such as deserts or mountainous regions, this approach is often severely constrained due to the absence of ground communication facilities \cite{10766404,11016724}. In UAV-assisted MEC systems, UAVs equipped with computing resources act as flying edge servers to support ground UEs in handling various computational tasks. By leveraging the mobility and flexibility of UAVs, such systems can significantly enhance the quality of user experience (QoE) \cite{10643301,10887202}.

Distinct from conventional ground-based cellular MEC systems and UAV-assisted MEC architectures, MEC systems based on LAENet typically involve multi-layered interactions between aerial and terrestrial components. Applications within the LAE framework not only include various ground devices but also commonly involve large aerial platforms that function as air taxis, balloons, and airborne charging stations. These platforms introduce additional fast-moving aerial users with inherent communication and computation demands. To meet the computational needs of such aerial users, high-altitude platforms (HAPs) can be equipped with edge servers and serve as high-altitude server nodes, providing computational support to a broad spectrum of both ground and aerial users \cite{9685072}. Deployed in the stratosphere, HAPs are unaffected by ground-level weather conditions and can offer stable communication and computation services across a wide area. In contrast, ground base stations are limited by their fixed locations, and UAVs are constrained by onboard resources and battery life, thus restricting their service coverage to localized regions. Therefore, in the MEC framework of LAENet, integrating the advantages of HAPs, ground base stations, and UAVs enables the provisioning of computational services for diverse ground and aerial users. This integrated architecture effectively expands the coverage of computing resources and enhances the resilience and flexibility of the network.

However, to fully leverage the advantages of the integrated MEC architecture within LAENet, several key challenges must be addressed. First, the offloading decisions among different user devices are interdependent, exhibiting coupling and high complexity. Moreover, the computation resource allocation strategies of MEC servers are closely intertwined with the offloading decisions made by user devices, further complicating the decision-making process. Second, the computing and communication resources of edge servers, such as HAPs and ground base stations, remain limited. Each user device must compete for these scarce resources. Therefore, under resource constraints, determining effective resource allocation strategies to meet the heterogeneous and stringent requirements of diverse tasks poses a significant challenge for MEC servers \cite{10382630}. Finally, due to the highly dynamic nature of LAENet, the locations, states, and task demands of various user devices are continuously changing. Additionally, environmental uncertainties can lead to instability in both aerial platforms and communication quality. As a result, developing task offloading and resource allocation strategies that are robust to complex task scenarios and responsive to unexpected conditions remains a critical challenge.

This paper investigates the problem of task offloading and resource allocation in MEC systems based on  LAENet. We aim to minimize the overall system cost regarding delay and energy consumption while satisfying offloading and resource constraints and ensuring performance guarantees. The main contributions of our work are summarized as follows.
\begin{itemize}
    \item \textcolor{black}{System Architecture: In the MEC system under LAENet, both aerial and ground users generate computation tasks and offload them via wireless connections. The offloaded tasks are processed collaboratively by high-altitude platforms and ground base stations. Specifically, the system comprises the following components: A high-altitude platform deployed in the upper layer provides wide-area communication and computation services. The low-altitude layer includes aerial devices such as UAVs and eVTOLs, which generate and execute computational tasks. The ground layer consists of ground users and ground base stations, where the latter provide more powerful but locally confined computation capabilities.} 
    \item \textcolor{black}{Problem Formulation: We model the low-altitude MEC system as a heterogeneous graph composed of multiple types of nodes and edges, each with associated features. Task offloading variables and resource allocation variables are defined on this graph structure. Based on this modeling, we formulate a novel joint optimization problem for task offloading and resource allocation, intending to minimize overall system delay and energy consumption. This optimization problem is further reformulated as a multi-task graph optimization problem for efficient solution.}
    \item \textcolor{black}{Algorithm Design: We propose graph attention diffusion-based solution generator (GADSG), a diffusion model algorithm built upon a Graph Attention Network (GAT) encoder as the backbone. The GAT encoder leverages multi-head attention mechanisms to extract information from the raw graph state and generate high-dimensional graph embeddings that encapsulate system environmental context and historical solution semantics. The diffusion model separately models the discrete offloading and continuous resource allocation variables. It constructs a forward diffusion process by injecting noise into the solution space. It employs a graph neural network to estimate the reverse denoising process, progressively generating optimized solutions in a high-dimensional latent space. The model is well-suited to the heterogeneous graph structure of LAENet, accommodating diverse node types and dynamically changing communication links.}
    \item Numerical Experiments: We conducted extensive experiments to evaluate the proposed algorithm. Compared to several baseline methods, GADSG demonstrates superior performance in optimization effectiveness, robustness, and generalization ability. This highlights the attention mechanism's capability to assign learnable weights to adjacent nodes and links, enabling dynamic modeling of critical connections in heterogeneous topologies.
\end{itemize}

The remainder of this paper is organized as follows. Section~\ref{RW} reviews the related work. Section~\ref{sec_problem} presents the system model and problem formulation. Section~\ref{GADSG} introduces the proposed graph attention-based diffusion method. Section~\ref{result} provides simulation results and performance analysis. Finally, Section~\ref{conclusion} concludes the paper.

\section{Related Work}\label{RW}
In this section, we review the related research on communication and computation in low-altitude economy networks, edge computing architectures, and network optimization. 

\subsection{Communication and Computation in LAENet}
As an emerging and integrated economic paradigm for future activities, the rapid development of the low-altitude economy has brought increasing attention to wireless communication and computing systems under LAENet. For example, \cite{10833672} investigates a resource scheduling problem for three-dimensional target search under the LAE scenario, where UAVs and ground vehicles collaboratively leverage edge computing. By applying a multi-agent deep reinforcement learning approach, the study optimizes task offloading and trajectory control to improve search efficiency and defend against intelligent jamming attacks. 
In \cite{sun2025taskdelayenergyconsumption}, an LAENet-based MEC system is studied, where a two-stage optimization method based on evolutionary multi-objective deep reinforcement learning is proposed to jointly optimize task offloading decisions and UAVs trajectories. In \cite{liu2025generativeailyapunovoptimization}, a novel approach integrating generative diffusion models with Lyapunov optimization theory is introduced to address system stability and performance optimization in UAV networks. 

Existing studies have made progress in areas such as network architecture, MEC systems, and communication robustness within LAENet, thereby contributing to the development of the low-altitude economy. However, most of these works are based on simplified assumptions and overlook the presence of heterogeneous aerial users and multi-layer cooperative MEC architectures, which are crucial for addressing the unique challenges posed by LAENet environments.

\subsection{Edge Computing System Architectures}
Edge computing, as a promising technology to address the limitations of computational resources and energy supply in user devices, has received significant attention and extensive research. For example, \cite{10124004} investigates a ground-based MEC network under uncertain system information and dynamic environments, where roadside vehicles randomly generate computation tasks that can be offloaded to ground base stations equipped with MEC servers to enhance user service satisfaction. In \cite{10225697}, a vehicular fog computing system is studied, in which task-generating vehicles can offload their tasks to service vehicles via vehicle-to-RSU (roadside unit) links, aiming to minimize the global average task delay. \cite{10134570} focuses on a multi-UAV cooperative computing network, where users offload tasks to computing-capable UAVs. By enabling cross-region UAV resource sharing and cooperative task processing, the system minimizes global task processing delay and balances energy consumption across UAVs. In \cite{10077418}, a vehicular edge computing (VEC) framework is proposed, where UAVs serve as mobile edge servers for vehicles and overloaded RSUs to minimize the average task processing delay. The authors in \cite{10914555} present a hierarchical aerial computing network, where UAVs and HAPs form a layered aerial computing architecture to compensate for the limitations of ground-based MEC. Ground users can offload tasks to UAVs, which must further decide whether to process the tasks locally or offload them to HAPs. In \cite{9978929}, a Low Earth Orbit (LEO) satellite edge computing system architecture is proposed, and the collaboration between LEO satellites and ground stations is investigated to provide efficient computational services. In \cite{10935306}, a three-tier space–air–ground integrated network architecture composed of LEO satellites, UAVs, and ground base stations is proposed, suitable for highly dynamic, multi-node cooperative scenarios.

In summary, existing edge computing architectures primarily focus on traditional ground-based MEC, vehicular fog computing, UAV-assisted MEC, and air–ground heterogeneous MEC systems. However, applications under the LAE framework also involve additional aerial users such as eVTOLs, which cannot independently execute computation-intensive and latency-sensitive tasks. Therefore, there is a need to design a novel LAENet-based MEC system that supports coordinated air–ground computing to address the limitations of existing research.

\begin{figure*}[t]
\centering
\includegraphics[width=5.25in]{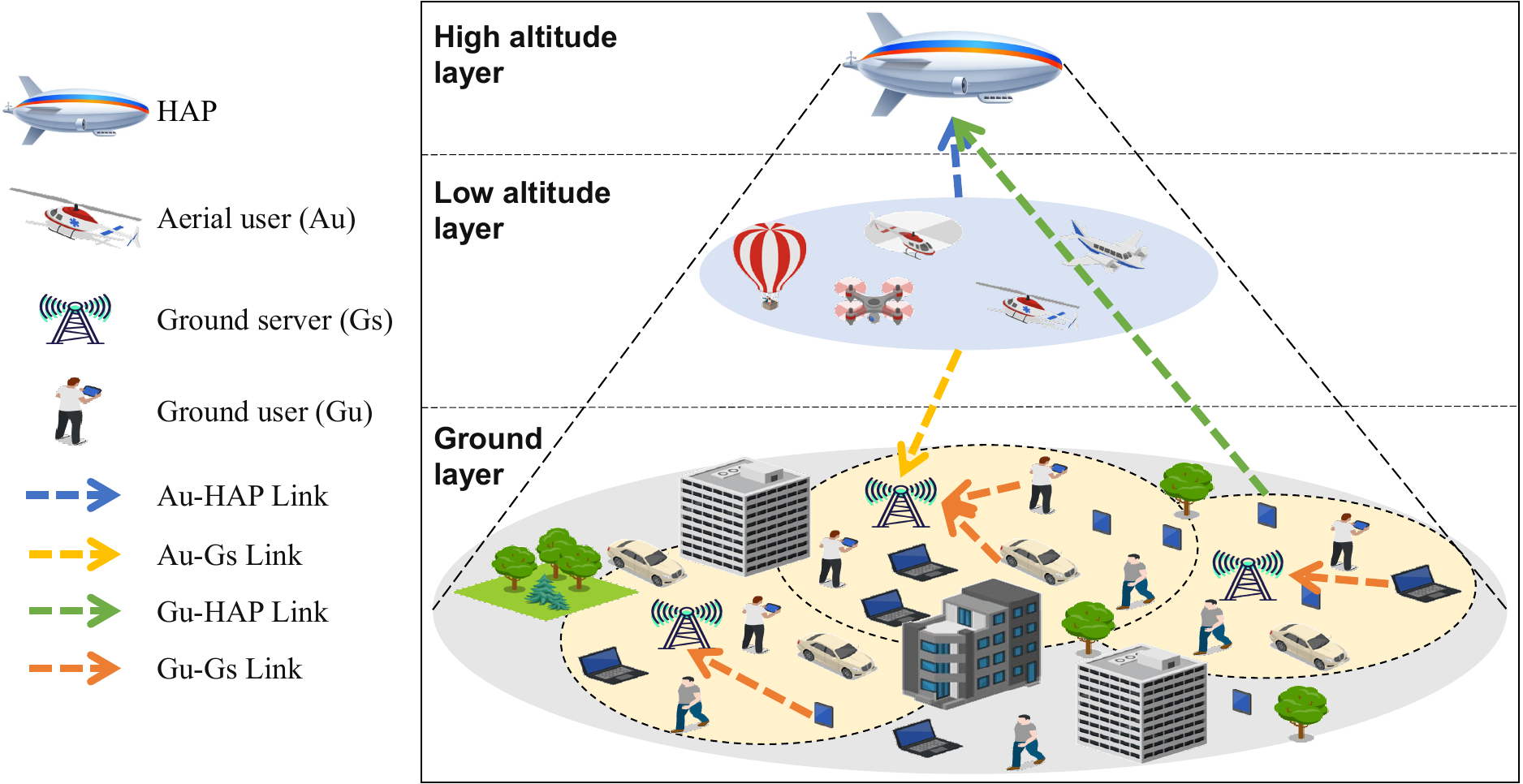}
\caption{The considered computation offloading model in LAENet-based MEC systems.
}
\label{fig_sys_model}
\vspace{-8pt}
\end{figure*}

\subsection{Network Optimization}
To address the complex optimization problems of task offloading and resource allocation, numerous researchers have focused on the design of efficient algorithms. Specifically, the most commonly adopted optimization approaches include numerical methods, swarm intelligence algorithms, and deep reinforcement learning (DRL) techniques.

For numerical methods and swarm intelligence algorithms, \cite{9305267} proposes an alternating iterative algorithm that employs block coordinate descent to optimize resource allocation and trajectory in separate stages. In the first stage, the trajectory is fixed, and resource allocation is optimized using the Lagrangian dual method. In the second stage, with fixed resource allocation, the trajectory is optimized using convex optimization techniques. \cite{9656594} proposes a near-optimal algorithmic framework based on semi-Lagrangian relaxation and distributed subgradient methods, which achieves joint optimization of task offloading, caching, and bandwidth resources via variable decomposition and iterative solving. \cite{9775682} introduces a Lyapunov optimization framework and combines it with a greedy strategy and a double-descent method to achieve online offloading optimization for long-term cost in dynamic multi-satellite–ground networks. In \cite{10382630}, a game-theoretic algorithm is used to optimize UAV offloading decisions, while a bisection method based on convex optimization and a genetic algorithm are proposed to determine the resource allocation ratio of edge servers. Regarding DRL-based methods, \cite{10887202} proposes a joint optimization algorithm that integrates matching theory with deep reinforcement learning to achieve multi-task cooperative optimization in UAV-assisted MEC systems. \cite{10833672} formulates the non-convex optimization problem with dynamic interference as a multi-agent problem, treating each user–server pair as an agent, and develops a multi-agent softmax deep deterministic policy gradient method for resource management. In \cite{10293163}, the optimization problem is reformulated as a Markov decision process, and a multi-agent deep deterministic policy gradient approach is used to enable low-complexity and real-time adaptive decision-making.

However, traditional numerical methods often require complex manual design and high computational complexity. Swarm intelligence algorithms exhibit low efficiency in high-dimensional spaces and typically incur long execution times. DRL methods, however, require carefully crafted reward and loss functions and depend on large numbers of training samples or frequent interactions with the environment to converge to an optimal policy.

In recent years, the rise of generative AI has provided a new methodology for network optimization. Particularly, generative diffusion models (GDMs), with their unique advantage in learning the intrinsic data distribution, have been shown to have significant potential for directly generating network resource optimization solutions \cite{11018297,10839314}. Thanks to their ability to generalize beyond the training data, GDMs have been verified to train on suboptimal data samples and generate high-quality solutions through parallel sampling \cite{11006143}. This characteristic effectively overcomes the traditional reliance of discriminative AI methods \cite{9079564} on high-cost optimal datasets. Recent research \cite{liang2025crossproblemsolvingnetworkoptimization} has proposed a problem-aware diffusion model (PAD), which enhances cross-problem generalization through a problem-aware learning framework. Furthermore, \cite{10736570} for the first time, diffusion models were combined with multi-agent reinforcement learning frameworks for optimizing task offloading decisions.

However, to date, less research has applied diffusion-based optimization algorithms to LAENet-based MEC systems, thereby failing to fully explore their potential. Given the highly dynamic nature of low-altitude airspace, real-time decision-making and robust processing capabilities become crucial. Therefore, we aim to design a robust algorithm capable of handling complex tasks, ensuring efficient and reliable task execution while optimizing system performance in a rapidly changing environment.

\section{System Model and Problem Formulation}\label{sec_problem}
In this section, we first present a hierarchical LAENet-based MEC architecture that incorporates aerial and ground users. Then, we introduce the corresponding system model and formulate the joint optimization problem. To effectively capture the heterogeneous and dynamic structure of the system, we abstract it as a graph and reformulate the objective as a multi-task graph optimization problem.
\vspace{-2.3mm}
\subsection{Overview of LAENet-based MEC Systems}
We consider a LAENet-based MEC framework as illustrated in Fig.~\ref{fig_sys_model}. The framework consists of three layers: a high-altitude layer composed of high-altitude platforms (HAPs), a low-altitude layer comprising multiple aerial users (AUs), and a ground layer consisting of multiple ground servers (GSs) and ground users (GUs). The HAPs are deployed in the stratosphere and powered by solar energy, enabling long-term operation. They provide wide-area wireless communication coverage and computing resources. The aerial users, including UAVs and eVTOLs, are characterized by high mobility, leading to dynamically changing network topologies. GSs offer dense coverage within a limited area, and GUs include devices such as smartphones and vehicles. AUs and GUs generate many computation-intensive tasks, but their local computing resources are limited. These tasks can be offloaded to HAPs or GSs for processing. The number of servers and users in the system is variable, allowing for dynamic node mobility and flexible node participation or departure based on task demands.

We formulate a binary computation offloading problem in LAENet-based MEC systems, aiming to minimize the weighted cost of system delay and energy consumption. The optimization variables include the offloading decision and the computational resource allocation of servers. As illustrated in Fig.~\ref{fig_sys_model}, the server set is defined as $\mathcal{S}\!=\!\{0,1,...,s,...,S\}$, where $s\!=\!0$ denotes the HAP acting as a server, and $s\!=\!\{1, ..., S\}$ represent GSs. Hence, there are $S\!+\!1$ servers providing computational resources. The set of AUs is denoted as $\mathcal{M}\!=\!\{0, 1, ..., m, ..., M\}$, and the set of GUs is denoted as $\mathcal{N}\!=\!\{0, 1, ..., n, ..., N\}$. Due to the limited radio coverage of each server, each user can only connect to a subset of the servers.

\subsection{Communication Model}
We consider four types of link channel models: the Ground User to Ground Station (G2G) channel model, the Ground User to HAP (G2H) channel model, the Aerial User to Ground Station (A2G) channel model, and the Aerial User to HAP (A2H) channel model. It is assumed that the uplink multiple access mechanism follows orthogonal multiple access (OMA) \cite{5779637} to ensure interference-free wireless transmission.

\subsubsection{G2G Channel Model}
In the LAENet-based MEC system, the communication links between GUs and GSs are primarily characterized by line-of-sight (LoS) propagation. According to \cite{10134570}, the channel power gain between GU $n$ and GS $s$ can be modeled as:
\begin{equation}
g_{n,s}^{G2G} = \frac{g_0}{l_{n,s}^2},
\label{g2g}
\end{equation}
where $g_0$ denotes the channel gain at a reference distance of 1 meter, $l_{n,s}$ is the distance between GU $n$ and GS $s$.

\subsubsection{G2H Channel Model}
Due to the large altitude difference between a GU $n$ and a HAP, the G2H communication link is primarily modeled as free-space propagation, with additional consideration of LoS and NLoS losses. According to \cite{10373079}, the path loss between GU $n$ and the HAP is modeled as:
\begin{equation}
L_n = 20 \log_{10} \left( \frac{4\pi f_c l_n}{c} \right) 
+ \rho_n^{\mathrm{LoS}} \eta_n^{\mathrm{LoS}} 
+ \left( 1 - \rho_n^{\mathrm{LoS}} \right) \eta_n^{\mathrm{NLoS}},
\end{equation}
where $l_{n}$ is the distance between GU $n$ and the HAP, $f_c$ is the carrier frequency, $c$ is the speed of light, and $\eta_{\mathrm{LoS}}$, $\eta_{\mathrm{NLoS}}$ denote the additional losses for LoS and NLoS links, respectively.

The LoS probability $\rho_n^{\mathrm{LoS}}$ can be calculated as \cite{10198895}:
\begin{equation}
\rho_n^{\mathrm{LoS}} = \frac{1}{1 + \kappa_n^1 \exp\left\{ -\kappa_n^2 \left[ \varphi_n - \kappa_n^1 \right] \right\}},
\end{equation}
where $\kappa_n^1$, $\kappa_n^2$ are environment-related parameters, and $\varphi_n$ is the elevation angle between GU $n$ and the HAP.

Without loss of generality, the channel gain between GU $n$ and the HAP is given by:
\begin{equation}
g_{n}^{G2H} = 10^{-\frac{L_{n}}{10}}.
\end{equation}

\subsubsection{A2G Channel Model}
For the A2G communication link, due to the presence of various obstacles and scatterers in real environments, the radio signal may suffer from shadowing or scattering. Therefore, we adopt a probabilistic path loss model that considers LoS and NLoS transmission probabilities and path losses. According to \cite{10914555}, the path loss between AU $m$ and GS $s$ is given by:
\begin{equation}
L_{m,s} = \rho_{m,s}^{\mathrm{LoS}} \cdot L_{m,s}^{\mathrm{LoS}} + (1 - \rho_{m,s}^{\mathrm{LoS}}) \cdot L_{m,s}^{\mathrm{NLoS}},
\end{equation}
where $L_{m,s}^{\mathrm{LoS}}$ and $L_{m,s}^{\mathrm{NLoS}}$ are defined as:
\begin{equation}
L_{\mathrm{LoS}} = 20 \log_{10} \left( \frac{4\pi f_c l_{m,s}}{c} \right) + \eta_{\mathrm{LoS}},
\end{equation}
\begin{equation}
L_{\mathrm{NLoS}} = 20 \log_{10} \left( \frac{4\pi f_c l_{m,s}}{c} \right) + \eta_{\mathrm{NLoS}}.
\end{equation}
Here, $l_{m,s}$ denotes the distance between AU $m$ and GS $s$, and $\eta_{\mathrm{LoS}}$, $\eta_{\mathrm{NLoS}}$ represent the excessive path losses under LoS and NLoS conditions, respectively.

The LoS probability between AU $m$ and GS $s$ is generally modeled as:
\begin{equation}
\rho_{m,s}^{\mathrm{LoS}} = \frac{1}{1 + \kappa_{m,s}^1 \exp\left\{ -\kappa_{m,s}^2 \left[\varphi_{m,s} - \kappa_{m,s}^1 \right] \right\}},
\end{equation}
where $\kappa_{m,s}^1$, $\kappa_{m,s}^2$ are environment-related parameters, and $\varphi_{m,s}$ is the elevation angle between AU $m$ and GS $s$.

Therefore, the channel gain between AU $m$ and GS $s$ is given by:
\begin{equation}
g_{m,s}^{A2G} = 10^{-\frac{L_{m,s}}{10}}.
\label{eq3_9}
\end{equation}

\subsubsection{A2H Channel Model}
For the A2H communication link, the transmission environment is typically unobstructed, and the communication is dominated LoS propagation. Therefore, the link between AU and a HAP can be approximately modeled as a free-space channel. The free-space path loss (FSPL) model \cite{8873672} is applied to describe the A2H communication. Thus, the path loss between AU $m$ and the HAP is given by:
\begin{equation}
L_m = 32.45 + 20 \log f_c + 20 \log d_m,
\end{equation}
where $d_{m}$ is the distance between AU $m$ and the HAP, $f_c$ is the carrier frequency, and $c$ is the speed of light.

Accordingly, the channel gain between AU $m$ and the HAP is given by:
\begin{equation}
g_{m}^{A2H} = 10^{-\frac{L_{m}}{10}}.
\label{eq3_11}
\end{equation}

\subsubsection{Communication Rate Model}
For all communication links, let $g$ denote the channel gain of a given link. According to Shannon's capacity formula, the transmission rate over the link is given by:

\begin{equation}
R = B \log_2\left(1 + \frac{P \cdot g}{\sigma^2}\right),
\label{eq3_12}
\end{equation}
where $B$ is the communication bandwidth of the link, $P$ is the transmission power of the user device on the link, and $\sigma^2$ is the noise power.

\subsection{Weighted Cost of Delay and Energy Consumption}
\subsubsection{Local Execution Cost}
In the system, let the data size of a user task be $D$ (in bits), and assume each bit of the task requires $a$ CPU cycles. Let the local computation frequency of the user device be $f^{l}$, and the chip's energy coefficient be $\kappa$. Then, the local computation delay is given by:
\begin{equation}
T^{\mathrm{loc}} = \frac{D \cdot a}{f^{l}}.
\end{equation}
The energy consumption for local computation depends on the user device's chip architecture \cite{10373079} and is expressed as:
\begin{equation}
E^{\mathrm{loc}} = \kappa \cdot \left(f^l\right)^3 \cdot T^{\mathrm{loc}}.
\end{equation}

To comprehensively consider the user's sensitivity to delay and energy consumption, let $w$ denote the weight of the task's delay cost. When delay is prioritized, $w \to 1$; when energy is prioritized, $w \to 0$. Therefore, the weighted cost of local execution is defined as:
\begin{equation}
J^{\mathrm{loc}} = w \cdot T^{\mathrm{loc}} + (1 - w) \cdot E^{\mathrm{loc}}.
\end{equation}

\subsubsection{Offloading Execution Cost}
In the system, let $F$ denote the total available computing frequency of a given server, and $R$ denote the communication rate of the link over which the user offloads the task. Let $P^{loc}$ be the user's transmission power, and the server operates at a fixed active power $P^{act}$. Then, the transmission delay for a task offloaded from the user to the server is given by:
\begin{equation}
T^{\mathrm{tr}} = \frac{D}{R}.
\end{equation}
The transmission energy consumption is expressed as:
\begin{equation}
E^{\mathrm{tr}} = P^{loc} \cdot T^{\mathrm{tr}}.
\end{equation}
Accordingly, the weighted transmission cost is:
\begin{equation}
J^{\mathrm{tr}} = w T^{\mathrm{tr}} + (1 - w) E^{\mathrm{tr}}.
\end{equation}
The full-load computation delay at the server is expressed as:
\begin{equation}
T^{\mathrm{exe}} = \frac{D \cdot a}{F}.
\end{equation}
The corresponding energy consumption under full-load execution is:
\begin{equation}
E^{\mathrm{exe}} = P^{act} \cdot T^{\mathrm{exe}}.
\end{equation}
Hence, the weighted execution cost at the server is given by:
\begin{equation}
J^{\mathrm{exe}} = w T^{\mathrm{exe}} + (1 - w) E^{\mathrm{exe}}.
\end{equation}

\subsubsection{Delay Constraint}
In this system, let $\tau$ denote the maximum tolerable delay for a given user task. Define $\lambda \in \{0,1\}$ as an indicator of whether local execution exceeds the delay threshold, and let $\mu \in [0,1]$ represent the minimum proportion of computing resources that must be allocated by the server to satisfy the delay requirement for offloaded execution. If the task cannot meet the delay constraint even under full-load execution, i.e., $T^{\mathrm{tr}} + T^{\mathrm{exe}} > \tau$, then $\mu = 0$; otherwise, $\mu$ is calculated as:
\begin{equation}
\mu = \frac{\alpha \cdot D}{(\tau - T^{\mathrm{tr}}) F}.
\end{equation}

\subsection{Graph-Based Modeling}
\subsubsection{Virtual Graph}
\begin{figure}[t]  
\centering
\includegraphics[width=3.0in]{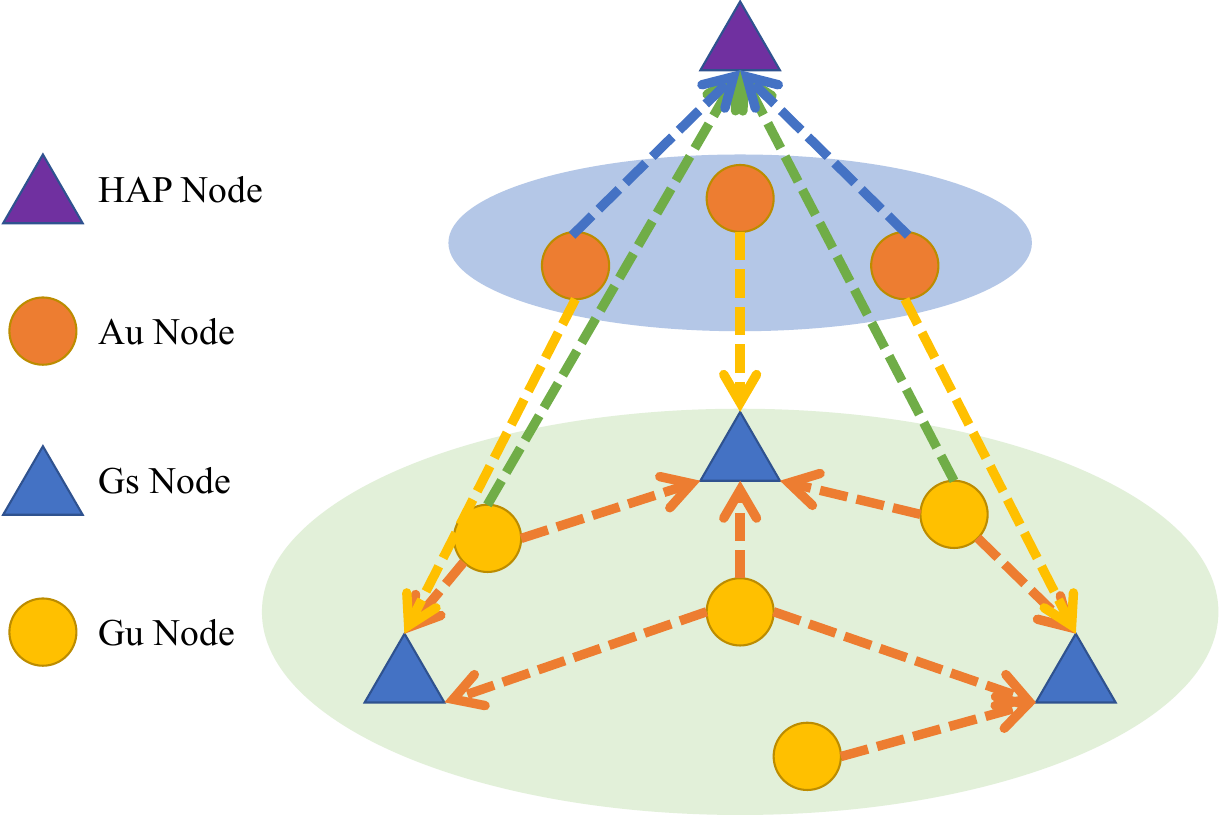}  
\caption{A toy example of the constructed virtual graph.}
\label{fig_vir_gra}
\vspace{-15pt}
\end{figure}

The problem is abstracted as a graph structure, resulting in the virtual graph shown in Fig.~\ref{fig_vir_gra}. Let the graph be denoted as $\mathcal{G}\!=\!(\mathcal{V}, \mathcal{E})$, where the node set is $\mathcal{V}\!=\!\mathcal{V}^M \cup \mathcal{V}^N \cup \mathcal{V}^S$, which includes all user nodes and server nodes. The edge set is defined as $\mathcal{E}\!=\!\{e_1, e_2,..., e_K\} \subseteq (\mathcal{V}^M \cup \mathcal{V}^N) \times \mathcal{V}^S$, where $|\mathcal{E}|\!=\!K$ denotes the total number of edges in the graph. Each edge $e_k\!=\!(\mathcal{V}^u, \mathcal{V}^s)$ represents a possible offloading link from AU or GU $\mathcal{V}^u \in \mathcal{V}^M \cup \mathcal{V}^N$ to a server $\mathcal{V}^s \in \mathcal{V}^S$.

Based on the previously established computation offloading model, in the virtual graph $\mathcal{G}$, each node $\mathcal{V}^j \in \mathcal{V}$ is associated with a feature vector $v_j^{fea}\!=\![\theta_j, D_j, f_j, w_j]$, where $\theta_j \in \{0, 1, 2, 3\}$ indicates the type of the node, corresponding to HAP, GS, AU, and GU, respectively. $D_j$ represents the input data size, $f_j$ is the computing frequency, and $w_j$ denotes the weight coefficient associated with delay sensitivity. For each edge $e_k$, the corresponding edge feature vector is defined as $e_k^{fea}\!=\! [j_k^{\mathrm{loc}}, j_k^{\mathrm{tr}}, j_k^{\mathrm{exe}}, \lambda_k, \mu_k]$, where $j_k^{\mathrm{loc}}$ denotes the weighted cost of local execution on the link, $j_k^{\mathrm{tr}}$ represents the transmission cost, and $j_k^{\mathrm{exe}}$ is the weighted cost of full-load execution on the server. In addition, $\lambda_k$ is a binary indicator of whether the local execution exceeds the delay constraint, and $\mu_k$ indicates the minimum proportion of computing resources required from the server for the offloaded task to meet the deadline.

\subsubsection{Optimization variables}
Let $X\!=\!(x_1, x_2, \dots, x_K)$ denote the offloading decision variables of users over all communication links in the system, where each element $x_k \in \{0, 1\}$ indicates whether the task is offloaded via edge $e_k$.To formalize the relationship between user nodes and the source ends of edges, we define the following binary function:
\begin{equation}\label{eq_source_node_j}
\begin{aligned}
    \Upsilon(e_k,\mathcal{V}^{i}) = \left\{
    \begin{matrix}
        1 & {\rm if }\ \mathcal{V}^{i} \ {\rm is\ the\ source\ node\ of }\ e_k \\ 
        0 & {\rm otherwise}
    \end{matrix}
    \right. \quad \\
    \forall k \in \{1, \dots, K\},\ \forall i \in \{1, \dots, M+N\}&.
\end{aligned}
\end{equation}
Since each user node $\mathcal{V}^i$ can offload its task to at most one server among all its outgoing edges, the system must satisfy the following constraint: $\sum_{\Upsilon(e_k,\mathcal{V}^{i})=1} x_k \leq 1$.

Let $Y\!=\!(y_1, y_2, \dots, y_K)$ denote the computing resource allocation ratio variables for each link in the system, where each element $y_k \in [0, 1]$ represents the proportion of computing resources allocated by the server to the user on edge $e_k$.To determine whether a server node $\mathcal{V}^s$ is the destination node of a directed edge $e_k$, we define the following binary function:
\begin{equation}\label{eq_des_node_j}
\begin{aligned}
    \Omega(e_k,\mathcal{V}^{s})=\left\{
    \begin{matrix}
        1 & {\rm if }\ \mathcal{V}^{s} \ {\rm is\ the\ destination\ node\ of }\ e_k \\ 
        0 & {\rm otherwise}
    \end{matrix}
    \right. \quad \\
        \forall k\in\{1,...,K\},\ \forall s\in\{0,...,S\}&.
\end{aligned}
\end{equation}
Since the total amount of computing resources allocated by a server node $\mathcal{V}^s$ to all tasks offloaded to it must not exceed its total available capacity, the system must satisfy the following constraint:$\sum_{\Omega(e_k,\mathcal{V}^{s})=1} x_k y_k \leq 1$.

Finally, it can be observed that the dimensions of both optimization variables $X$ and $Y$ are equal to the number of edges in the graph. Moreover, the variables $x_k$ and $y_k$ can be abstracted as the indicators of the existence and the weights of the corresponding edges in the graph, respectively.

\subsection{Weighted Cost Minimization Problem}
Based on the definitions of node features, edge features, and optimization variables, the weighted cost minimization problem of delay and energy consumption in the MEC system can be formulated as follows:
\begin{subequations}\label{eq_obj_func}
\begin{align*}
    &  \mathbb{P}: \underset{\{\mathcal{X},\mathcal{Y}\}}{{\rm min}}\ \sum^K_{k=1}(1-x_k)J^{\mathrm{loc}}_k+x_k(J^{\mathrm{tr}}_k+\frac{J^{\mathrm{exe}}_k}{y_k}), \\
    &s.t.\ \mathbf{C1}: \ x_k\in\{0,1\}, \ \forall k \in \{1,...,K\}, \\
    &\ \ \ \ \ \mathbf{C2}: \ y_k\in[0,1], \ \forall k \in \{1,...,K\}, \\
    &\ \ \ \ \ \mathbf{C3}: \ \sum_{\Upsilon(e_k,\mathcal{V}^{i})=1} x_k\leq 1, \forall i\in\{1,...,M+N\}, \\
    &\ \ \ \ \ \mathbf{C4}: \ \sum_{\Omega(e_k,\mathcal{V}^{s})=1} x_k y_k\leq 1, \forall s\in\{0,...,S\}, \\
    &\ \ \ \ \ \mathbf{C5}: \ (1-x_k)\lambda_k+\mu_kx_k>0.
\end{align*}
\end{subequations}

Constraint $\mathbf{C1}$ restricts the offloading decision variable $x_k$ to be binary. Constraint $\mathbf{C2}$ ensures that the computing resource allocation ratio $y_k$ for each user is positive and does not exceed the server's maximum available resources. Constraint $\mathbf{C3}$ guarantees that each user can offload its task to at most one server. Constraint $\mathbf{C4}$ ensures that the total computing resources allocated by a server to all offloaded tasks do not exceed its available capacity. Constraint $\mathbf{C5}$ guarantees that the total completion time of each task does not exceed the maximum tolerable delay. This problem involves both binary variables (offloading decisions) and continuous variables (computing resource allocation ratios), making it a non-convex and NP-hard Mixed Integer Nonlinear Programming (MINLP) problem.

Let $(\mathcal{X}^*, \mathcal{Y}^*)$ be the optimal solution given the input parameters of the graph structure. Problem $\mathbb{P}$ can then be transformed into a multi-task graph optimization problem \cite{11006143}. By utilizing the node type information in the node features and the objective/constraint-related information in the edge features, the graph-based formulation provides rich structural information. It can be extended to similar types of problems. In the following section, we propose a graph diffusion model-based approach to solve this graph optimization problem.

\section{The Proposed GADSG Algorithm} \label{GADSG}
This section introduces the proposed GADSG optimization algorithm. First, the overall framework of GADSG is outlined. Then, the encoder design based on the Graph Attention Network (GAT) is presented. Finally, the solution process of obtaining the optimal offloading decisions and resource allocation ratios using a graph diffusion model is described.

\subsection{Framework Overview}
\begin{figure*}[t]
\centering
\centerline{\includegraphics[width=5.8in]{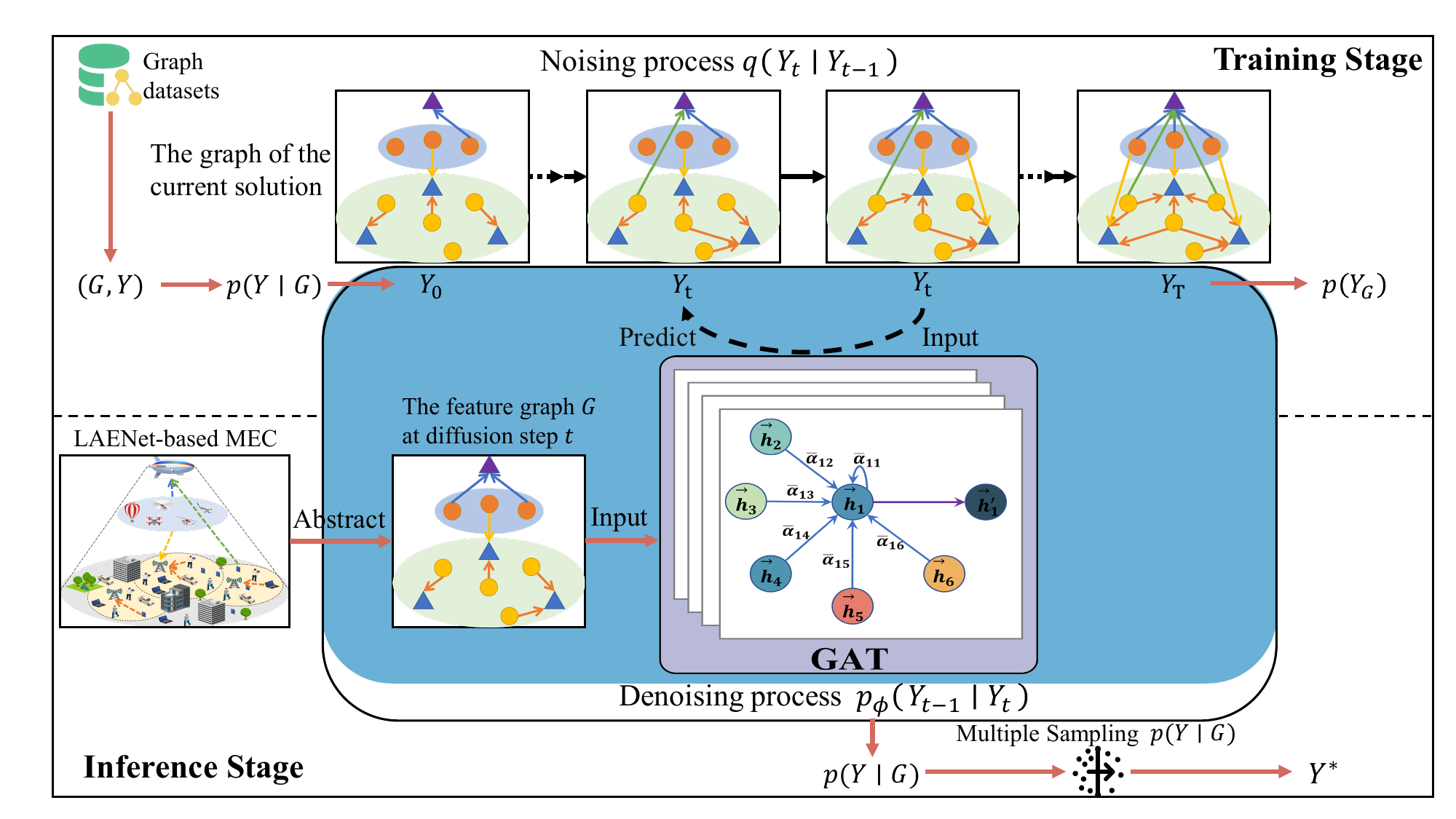}}
\caption{The algorithmic framework of the proposed GADSG method.}
\label{fig_sol_gen}
\vspace{-0.6cm}
\end{figure*}

The diffusion model algorithm proposed in this work is structured around a GAT-based encoder backbone, as illustrated in Fig.~\ref{fig_sol_gen}. Specifically, at each time step $t$, the system is modeled as a heterogeneous graph, where nodes represent different types of users and servers, and edges denote possible offloading links, augmented with edge features such as channel gain and execution cost. This graph, together with the current solution state, is fed into the GAT encoder for feature extraction. The encoder sequentially performs edge padding, gated filtering, feature embedding, and multi-head attention aggregation to generate high-dimensional graph embedding features, which encode both system environment information and the semantic representation of historical solutions. These embeddings serve as the initial conditions for the subsequent diffusion steps. In the graph diffusion module, the offloading decisions (discrete variables) and resource allocation ratios (continuous variables) are modeled through separate diffusion processes. A forward diffusion process is constructed by injecting noise into the solution space, and a reverse denoising process is learned using graph neural networks to generate optimized solutions in the high-dimensional latent space iteratively. During training, the negative log-likelihood variational lower bound is adopted as the joint optimization objective. A unified loss is backpropagated through both solution generation processes to enhance the collaborative optimization capability of the model. After training, the model can directly take the current system graph as input and rapidly sample high-quality offloading and resource allocation solutions that satisfy system constraints.

\subsection{GAT-Based Encoder}
As described in Sec. \ref{sec_problem}, we abstract the LAENet-based MEC system as a heterogeneous graph containing four types of nodes. Each edge carries multi-dimensional attributes such as channel gain and execution cost, and both node and edge features dynamically evolve with the mobility of user devices. Due to the heterogeneity and dynamic nature of the network graph, conventional convolutional GNNs struggle to distinguish critical offloading links from less relevant connections. To address these challenges, we propose a GAT-based encoder that supports sparse and dense implementations, making it adaptable to rapid changes in node scale and network topology.

Specifically, the GAT encoder takes as input the current diffusion time step $t$, the feature graph $\mathcal{G}$, and the current offloading decision and resource allocation solution, which shares the same adjacency structure as $\mathcal{G}$. The feature graph and the current solution are first merged, and missing edges are completed through an edge padding mechanism to construct a homogeneous fully-connected graph. Then, unified linear embeddings are applied to both node and edge features, followed by the addition of 2D positional encoding and temporal embedding. Within each GNN layer, the encoder filters out padded virtual edges using a gating mechanism. It performs parallel multi-head graph attention over the remaining valid neighborhood to learn neighbor importance in different semantic subspaces and conduct soft weighted aggregation. Finally, recursive message passing is performed across GNN layers to produce high-quality input features for the next diffusion time step.

\subsubsection{Edge Padding and Gating Mechanism}

\begin{itemize}
    \item \textbf{Edge padding and feature embedding:}
In LAE-based MEC scenarios, the coverage of each edge server is limited, and user nodes are not fully connected to server nodes. Moreover, the user–server topology dynamically changes with user mobility. As a result, the original graph is sparsely connected, and the adjacency matrices vary in size across different samples. To ensure that all samples maintain a consistent homogeneous dimension during batch training, we first perform edge padding for each sample. Specifically, for any edge $e_K$, a binary existence flag is inserted into its original 5-dimensional feature vector to construct a new edge feature vector:
\begin{equation}
\mathbf{e}_k^{\mathrm{Pad}} = \left[ \underset{\text{\scriptsize flag}}{\underbrace{1}},\ J_k^{\mathrm{loc}},\ J_k^{\mathrm{tr}},\ J_k^{\mathrm{exe}},\ \lambda_k,\ \mu_k \right].
\end{equation}
Since there are a total of $S + 1$ servers in the system, the maximum number of servers that each user can connect to is $S + 1$. If the actual number of connected edges is less than this upper bound, virtual edges are added to complete the missing connections, and these virtual edges are assigned with penalized feature values:
\begin{equation}
\mathbf{e}_{k'}^{\mathrm{Pad}} = \left[ \underset{\text{\scriptsize flag}}{\underbrace{0}},\ J_{k}^{\mathrm{loc\text{-}m}},\ J_{k}^{\mathrm{tr\text{-}m}},\ J_{k}^{\mathrm{exe\text{-}m}},\ 0,\ 0 \right],
\end{equation}
where $j_k^{(\mathrm{loc}\text{-}m)}$, $j_k^{(\mathrm{tr}\text{-}m)}$, and $j_k^{(\mathrm{exe}\text{-}m)}$ denote the maximum observed costs within the dataset, serving to penalize the selection of virtual (nonexistent) links in the loss function. The two zero vectors are included merely for dimensional alignment and do not contribute to gradient backpropagation.

After edge padding is completed, the encoder feeds the node and edge features into an embedding layer, which is projected into a unified hidden dimension $d_h$. To capture the spatiotemporal dynamics of the network, a 2D sinusoidal positional encoding is added to each node, and a time embedding related to the current diffusion step is injected into the entire graph.

\item \textbf{Gating mechanism}: 
To prevent virtual edges from participating equally with real edges during message propagation, thus severely diluting meaningful information, we introduce a differentiable gating mechanism into each GNN layer to dynamically regulate information flow. Specifically, for each edge $e_{us}$, the model generates a gating weight based on the paired node features and edge features:
\begin{equation}
\sigma_k = \mathrm{sigmoid}\left( W_1 \mathbf{h}_u + W_2 \mathbf{h}_s + W_3 \mathbf{e}_{k}^{\mathrm{fea}} \right) \cdot \mathrm{flag}_{k'},
\end{equation}
where $W_1$, $W_2$, and $W_3$ are learnable parameters, $e_{us}^{\mathrm{fea}}$ denotes the edge feature embedding for the edge from source user node $u$ to target server node $s$, $h_u$ and $h_s$ are the feature embeddings of the corresponding nodes, and $\mathrm{sigmoid}$ is the activation function. $\mathrm{flag}_k$ refers to the previously inserted edge existence indicator. As a result, information is propagated from $u$ to $s$ only when $\sigma_k > 0$.
\end{itemize}

\subsubsection{Multi-Head Graph Attention-Based Message Aggregation}
GATs incorporate the self-attention mechanism into graph structures, enabling the model to adaptively evaluate and assign weights to neighboring nodes during message propagation~\cite{veličković2018graphattentionnetworks}. This allows the network to highlight the most critical connections for offloading decisions while suppressing redundant or noisy information. In addition, the attention mechanism naturally supports variable-length inputs, making it highly suitable for computation offloading scenarios where the number of nodes dynamically changes.

After edge padding and gated filtering are completed, it remains necessary to further distinguish the importance of adjacent server nodes for each user node to improve the accuracy of task offloading and resource allocation decisions. To this end, we introduce a multi-head graph attention mechanism to perform weighted aggregation, enabling the model to select more optimal offloading decisions.

\begin{itemize}
    \item \textbf{Single-head attention scoring and weight normalization}
For a real edge $e_{us}$ with source user node $u$ and target server node $s$, the attention score of the $h$-th attention head is first computed as:
\begin{equation}
e_{us}^{(h)} = \vec{a}^{(h)} \cdot \tanh\left( W \mathbf{h}_u + W \mathbf{h}_s \right),
\end{equation}
where $\vec{a}^{(h)}$ is the learnable attention vector for the $h$-th head, and $W$ is the shared linear transformation weight. To further incorporate the influence of edge attributes into attention computation, a gating-based correction term derived from edge features is introduced within each attention head:
\begin{equation}
g_{us}^{(h)} = \mathrm{sigmoid}\left( \mathbf{W}_g^{(h)} \cdot \mathbf{e}_{us} \right),
\end{equation}
where $W_g^{(h)}$ is a learnable parameter matrix. Finally, this gating factor is combined with the attention score to compute the normalized attention weight.
\begin{equation}
\tilde{\alpha}_{us}^{(h)} = 
\frac{
\exp\left( e_{us}^{(h)} \cdot g_{us}^{(h)} \cdot \mathrm{flag}_{us} \right)
}{
\sum\limits_{k \in \mathcal{N}(u)} \exp\left( e_{uk}^{(h)} \cdot g_{uk}^{(h)} \cdot \mathrm{flag}_{uk} \right) + \epsilon
},
\end{equation}
where $\mathcal{N}(u)$ denotes the set of neighboring nodes of $u$, and $\epsilon$ is a small constant added for numerical stability to avoid division by zero.
\item \textbf{Multi-head attention aggregation and feature update}
We adopt a multi-head attention mechanism to capture neighbor relations at different semantic levels. Each head performs attention computation in an independent subspace, focusing on different types of information and relational patterns. The outputs of the $H$ attention heads are then concatenated or averaged to generate the final neighborhood representation:
\begin{equation}
\mathbf{h}_u' = \mathrm{Concat}_{h=1}^{H} \left( \sum_{j \in \mathcal{N}(u)} \tilde{\alpha}_{uj}^{(h)} \cdot \mathbf{h}_j \right),
\end{equation}
where $\mathrm{Concat}_{h=1}^H(\cdot)$ denotes the concatenation of outputs from all $H$ attention heads.

Finally, the node features are updated as:
\begin{equation}
h_u^{\mathrm{new}} = \mathrm{ReLU}(h_u + W_{\mathrm{attn}} h_u'),
\end{equation}
where $W_{\mathrm{attn}}$ is a learnable linear transformation matrix. Edge features are updated synchronously:
\begin{equation}
e_{us}^{\mathrm{new}} = \mathrm{ReLU}(A h_u + B h_s + C e_{us}),
\end{equation}
so that edge features are coupled with node states in a closed loop, allowing the iterative process to converge toward a weight representation that reflects the current state of resource competition.
\end{itemize}

\subsection{GADSG Optimization Algorithm}
Diffusion models, which simulate the gradual recovery process from noise to structured solution space~\cite{NEURIPS2020_4c5bcfec}, can be utilized to generate high-quality optimization solutions. In the forward diffusion process, the model progressively corrupts the original solution by injecting noise, driving it toward a random distribution. Then, in the reverse denoising phase, the model gradually removes the noise to recover a solution close to the optimum. This process supports generating continuous variables and discrete decisions, making it well-suited for complex multi-dimensional optimization problems such as task offloading and resource allocation. Moreover, because diffusion models generate solutions by learning the solution space distribution conditioned on the input, they have been shown to possess stronger generalization capabilities than discriminative models in complex network optimization tasks~\cite{11018297,10839314,11006143}. In this section, we propose the GADSG optimization algorithm to solve the graph-based task offloading and resource allocation problem formulated in Sec. \ref {sec_problem}.

Specifically, the input to the graph diffusion model comes from the node and edge embeddings generated by the GAT encoder, which represent the initial features of the network topology, task offloading, and resource allocation. The diffusion model optimizes both the discrete offloading decision and the continuous resource allocation through a forward noise injection process and a reverse denoising process. At each diffusion step, a graph neural network aggregates structural information from the graph to guide the recovery of the optimal solution, ensuring that the generated offloading and resource allocation schemes are feasible and efficient under the constraints of the graph structure. In the following, for the input graph $\mathcal{G} = (\mathcal{V}, \mathcal{E})$, we detail the generation processes of the discrete offloading decision solution $X^*$ and the continuous resource allocation solution $Y^*$, respectively.

\subsubsection{Diffusion Generation of Offloading Decisions}
\begin{itemize}
    \item \textbf{Forward noise injection process}:
For any feasible user–server edge in the graph (i.e., an edge over which offloading may occur), the offloading decision variable is denoted as $x \in \{0,1\}$. We first parameterize the complete set of offloading decisions $X$ following the method in~\cite{11006143} as a matrix-form one-hot vector $X \in \{0,1\}^{K \times 2}$, where $K$ is the number of feasible edges. The forward diffusion process gradually corrupts the original solution $X_0$ into a sequence of noisy representations $X_1, X_2, \dots, X_T$ by introducing a series of transition matrices $Q_t \in \mathbb{R}^{2 \times 2}$. Each $Q_t$ is defined as a symmetric matrix:
$Q_t = \begin{bmatrix} 1 - \beta_t & \beta_t \\ \beta_t & 1 - \beta_t \end{bmatrix}$, where $\beta_t \in (0,1)$ controls the noise intensity at diffusion step $t$, and it satisfies the condition $\prod_{t=1}^T (1 - \beta_t) \approx 0$. Based on this, the single-step and $t$-step diffusion distributions are given by:
\begin{equation}
q(X_t \mid X_{t-1}) = \mathrm{Cat}(X_t;\ \mathbf{p} = X_{t-1} Q),
\end{equation}
\begin{equation}
q(X_t \mid X_0) = \mathrm{Cat}(X_t;\ \mathbf{p} = X_0 \bar{Q}_t),
\end{equation}
where $\mathrm{Cat}(\cdot)$ denotes a categorical distribution, which can be any arbitrary discrete distribution. The cumulative transition matrix is $\bar{Q}_t = Q_1 Q_2 \dots Q_t$. This distribution indicates that each $X_t$ is obtained by multiplying the previous state $X_{t-1}$ with the corresponding transition matrix, ultimately ensuring that $X_T \sim \mathrm{Uniform}(\cdot)$.

\item \textbf{Reverse denoising process:}
In the reverse denoising process, the model starts from $X_T \sim \mathrm{Uniform}(\cdot)$ and progressively removes noise by leveraging the graph structural information to recover the target decision $X_0$. The denoising process is modeled as a parameterized conditional Markov chain:
\begin{equation}
p_\Phi(X_{t-1} \!\mid\! X_t, G) \!=\! \sum_{\tilde{X}_0} q\left(X_{t-1} \!\mid\! X_t, \tilde{X}_0 \right) \cdot p_\Phi\left( \tilde{X}_0 \!\mid\! X_t, G \right),
\end{equation}
where $p_{\Phi}(\tilde{X}_0 \mid X_t, \mathcal{G})$ denotes the probability distribution of the original solution predicted by a graph neural network, and $\mathcal{G}$ represents the input task offloading graph composed of node and edge features. According to Bayes’ theorem, the posterior distribution is computed as:
\begin{equation}
q(X_{t-1} \!\mid\! X_t, X_0) \!=\! \mathrm{Cat} \left(X_{t-1};\ \mathbf{p} \!=\! 
\frac{X_t Q_t^\top \odot X_0 Q_{t-1}}{X_0 Q_t X_t^\top} \right),
\end{equation}
where $\odot$ denotes element-wise multiplication. 

We can replace $X_0$ in the equation with the estimated value $\tilde{X}_0$ predicted by the GNN to complete the sampling process.
\end{itemize}

\subsubsection{Diffusion Generation of Resource Allocation}
\begin{itemize}
    \item \textbf{Forward noise injection process:}
For any feasible user–server edge in the graph, the resource allocation solution is defined as $y \in [0,1]$. Therefore, the overall resource allocation solution can be represented as a vector $Y \in \mathbb{R}^K$, where $K$ is the number of feasible edges. The diffusion process injects standard Gaussian noise step-by-step to perturb the original solution into a near-random state progressively. The distributions for the single-step and $t$-step continuous diffusion processes are given by:
\begin{equation}
q(Y_t \mid Y_{t-1}) = \mathcal{N}\left(Y_t;\ \sqrt{1 - \beta_t} Y_{t-1},\ \beta_t \mathbf{I} \right),
\end{equation}
\begin{equation}
q(Y_t \mid Y_0) = \mathcal{N}\left(Y_t;\ \sqrt{\bar{\alpha}_t} Y_0,\ (1 - \bar{\alpha}_t) \mathbf{I} \right),
\end{equation}
where $\beta_t \in (0,1)$ controls the noise intensity at step $t$, and $\bar{\alpha}_t = \prod_{s=1}^t (1 - \beta_s)$ denotes the cumulative noise factor. The condition $\prod_{t=1}^T (1 - \beta_t) \approx 0$ ensures that $Y_T \sim \mathcal{N}(\cdot)$.

    \item \textbf{Reverse denoising process:}
In the reverse denoising process, the model starts from $Y_T \sim \mathcal{N}(\cdot)$ and gradually removes noise by incorporating graph structural information to recover the target solution $Y_0$. At each step $t$, the model uses a graph neural network to predict the noise component contained in the current solution, denoted as $\epsilon_\phi(Y_t, t, \mathcal{G})$, and then estimates the original solution $\tilde{Y}_0$ based on this noise:
\begin{equation}
\tilde{Y}_0 = \frac{1}{\sqrt{\bar{\alpha}_t}} \left( Y_t - \sqrt{1 - \bar{\alpha}_t} \cdot \epsilon_\phi(Y_t, t, G) \right).
\end{equation}
Subsequently, the estimated solution is used to construct the mean term of the Gaussian noise, based on which the reverse denoising distribution at each step is defined as:
\begin{equation}
p_\phi(Y_{t-1} \mid Y_t) = \mathcal{N}\left( Y_{t-1};\ \mu_\phi(Y_t, t),\ \sigma_t^2 \mathbf{I} \right),
\end{equation}
where the mean term $\mu_\phi(Y_t, t)$ is defined as:
\begin{equation}
\mu_\phi(Y_t, t) = \frac{1}{\sqrt{\alpha_t}} \left( 
Y_t - \frac{1 - \alpha_t}{\sqrt{1 - \bar{\alpha}_t}} \cdot \epsilon_\phi(Y_t, t, G) 
\right).
\end{equation}
\end{itemize}

\begin{algorithm}
\caption{Hybrid Diffusion Training}
\label{alg:hybrid_diffusion}
\KwIn{Training dataset $\mathcal{D} = \{(G, y_{\mathrm{cls}}, y_{\mathrm{reg}})\}$; diffusion steps $T$; noise schedule $\{\alpha_t\}_{t=1}^{T}$; model parameters $\theta$;}
\KwOut{Trained model parameters $\theta$;}

\For{each training step}{
  Sample $(G, y_{\mathrm{cls}}, y_{\mathrm{reg}}) \sim \mathcal{D}$\;
  Sample time step $t \sim \mathrm{Uniform}(\{1, \dots, T\})$\;
  Sample Gaussian noise $\boldsymbol{\epsilon} \sim \mathcal{N}(0, \mathbf{I})$\;
  Compute noised input: $y_{\mathrm{reg}}^{(t)} = \sqrt{\bar{\alpha}_t} \cdot y_{\mathrm{reg}} + \sqrt{1 - \bar{\alpha}_t} \cdot \boldsymbol{\epsilon}$\;
  One-hot encode discrete label: $y_{\mathrm{cls}}^{(t)} = \mathrm{OneHot}(y_{\mathrm{cls}})$\;
  Concatenate features: $x_t = [y_{\mathrm{cls}}^{(t)},\ y_{\mathrm{reg}}^{(t)}]$\;
  Predict noise via GNN encoder: $\hat{\boldsymbol{\epsilon}} = \boldsymbol{\epsilon}_\theta(x_t, t, G)$\;
  Separate predictions: $\hat{\boldsymbol{\epsilon}}_{\mathrm{cls}} = \hat{\boldsymbol{\epsilon}}[:, :2]$, $\hat{\boldsymbol{\epsilon}}_{\mathrm{reg}} = \hat{\boldsymbol{\epsilon}}[:, 2]$\;
  Compute loss: $\mathcal{L} = \mathrm{CrossEntropy}(\hat{\boldsymbol{\epsilon}}_{\mathrm{cls}}, y_{\mathrm{cls}}) + \|\hat{\boldsymbol{\epsilon}}_{\mathrm{reg}} - \boldsymbol{\epsilon}\|_2^2$\;
  Update: $\theta \leftarrow \theta - \eta \cdot \nabla_\theta \mathcal{L}$\;
}
\textbf{end for}
\end{algorithm}

\subsubsection{Model Training and Inference}
To optimize the solution generation process, the diffusion model is trained to approximate the true distribution of offloading decisions by minimizing a loss function, as illustrated in \textbf{Algorithm~\ref{alg:hybrid_diffusion}}. For both discrete and continuous solution generation, the training objective is defined as the negative log-likelihood variational upper bound:
\begin{align}
\mathcal{L} \!&=\! \mathbb{E}_q \Big[ 
 \!-\! \log p_\Phi(Y_0 \mid Y_1, G) \notag \\
& \!+\! \sum_{t=2}^{T} D_{\mathrm{KL}} \big( 
q(Y_{t-1} \mid Y_t, Y_0)\,\|\,p_\Phi(Y_{t-1} \mid Y_t, G)
\big) 
\Big] \!+\! C,
\end{align}
where $D_{\mathrm{KL}}(\cdot \,\|\, \cdot)$ denotes the Kullback–Leibler (KL) divergence, which measures the discrepancy between the generative distribution and the target distribution. The constant term $C$ includes regularization components to prevent overfitting.

Specifically, the generation of offloading decisions is optimized using the cross-entropy loss function:
\begin{equation}
\mathcal{L}_{\mathrm{disc}} = \mathrm{CE}\left( X_{\mathrm{cls}}^{\mathrm{pred}},\ X_{\mathrm{cls}}^{\mathrm{true}} \right),
\end{equation}
where $X_{\text{cls}}^{\text{pred}}$ and $X_{\text{cls}}^{\text{true}}$ denote the predicted and ground-truth offloading decisions, respectively, and $\mathrm{CE}(\cdot)$ represents the cross-entropy function. For the generation of resource allocation solutions, an additive Gaussian noise model is adopted, and the mean squared error (MSE) loss function is used:
\begin{equation}
\mathcal{L}_{\mathrm{cont}} = \left\| \epsilon - \epsilon_\phi(Y_t, t, G) \right\|^2,
\end{equation}
where $\epsilon$ denotes the ground-truth noise in the diffusion process, and $\epsilon_\phi(Y_t, t, \mathcal{G})$ represents the noise estimation predicted by the model.

The graph diffusion model has been shown to satisfy the inter-task gradient orthogonality principle for both discrete and continuous diffusion tasks~\cite{11006143}. To enable end-to-end joint training for the dual-task objective, we adopt a unified network architecture during training, where both the offloading decisions and resource allocation ratios are predicted along the same diffusion trajectory. The joint loss is defined as:
\begin{equation}
\mathcal{L}_{\mathrm{total}} = \mathcal{L}_{\mathrm{disc}} + \mathcal{L}_{\mathrm{cont}}.
\end{equation}

\begin{algorithm}[t]
\caption{Hybrid Diffusion Sampling}
\label{alg:hybrid_sampling}
\KwIn{Trained model parameters $\theta$; diffusion steps $T$; noise schedule $\{\alpha_t\}_{t=1}^{T}$; input graph $G$;}
\KwOut{Discrete offloading decision $\hat{y}_{\mathrm{cls}}$, continuous resource allocation $\hat{y}_{\mathrm{reg}}$;}

Initialize: $y_T \sim \mathcal{N}(0, \mathbf{I})$\;

\For{$t = T, T{-}1, \dots, 1$}{
  \eIf{$t > 1$}{
    Sample $\boldsymbol{\epsilon} \sim \mathcal{N}(0, \mathbf{I})$\;
  }{
    Set $\boldsymbol{\epsilon} = 0$\;
  }
  Predict noise: $\hat{\boldsymbol{\epsilon}} = \boldsymbol{\epsilon}_\theta(y_t, t, G)$\;
  Compute the denoised result using the DDPM update rule:
  \begin{align*}
    y_{t-1} \!=\! \frac{1}{\sqrt{\alpha_t}} \bigg( y_t \!-\! \frac{1 \!-\! \alpha_t}{\sqrt{1 \!-\! \bar{\alpha}_t}} \cdot \hat{\boldsymbol{\epsilon}} \bigg)
    \!+\! \sqrt{1 \!-\! \bar{\alpha}_{t-1}} \cdot \boldsymbol{\epsilon}
  \end{align*}
}
\textbf{end for}
Extract outputs: $\hat{y}_{\mathrm{cls}},\ \hat{y}_{\mathrm{reg}}$\;

\Return{$(\hat{y}_{\mathrm{cls}},\ \hat{y}_{\mathrm{reg}})$}
\end{algorithm}

Finally, we incorporate the DDIM framework as a sampling accelerator~\cite{song2022denoisingdiffusionimplicitmodels} to enable sampling. The pseudocode of the sampling procedure is summarized in \textbf{Algorithm~\ref{alg:hybrid_sampling}}. This mechanism performs non-Markovian sampling by skipping intermediate timesteps and directly sampling key diffusion steps, significantly reducing inference overhead. Specifically, the inference process starts from a Gaussian prior state and skips over intermediate steps according to a predefined stride schedule. The graph neural network is only executed at selected key steps to perform prediction and decoding, resulting in high-quality offloading and resource allocation solutions. 

\section{Experiments} \label{result}
In this section, we construct multiple instances of LAENet-based MEC problems at various scales to validate the effectiveness of the proposed GADSG model. Several algorithms are selected as baselines, and comprehensive comparisons are conducted regarding convergence, optimization performance, robustness, and generalization capability.
\subsection{Simulation Setup and Dataset Construction}
\subsubsection{Simulation Setup}
In this study, we build a representative simulation environment for a multi-user, multi-server collaborative LAENet-based MEC system, aiming to emulate realistic decision-making scenarios for task offloading and resource allocation. As shown in Table~\ref{tab:hyperparams}, we define the range of MEC-related hyperparameters based on statistical data from the prototype scenario, including the computational demands of user tasks, device processing capabilities, and communication parameters.

\begin{table}[t]
\centering
\caption{Hyperparameters of MEC system}
\label{tab:hyperparams}
\renewcommand{\arraystretch}{1.25}
\resizebox{\linewidth}{!}{%
\begin{tabular}{|c|c|}
\hline
\textbf{Hyperparameter} & \textbf{Value / Distribution} \\
\hline
Computation workload $D_l$ [FLOPs] & 
\makecell[l]{Norm($\mu = 4.5\mathrm{e}{9}$, $\sigma = 2.5\mathrm{e}{9}$,\\ \quad $[1.2\mathrm{e}{9}, 7\mathrm{e}{9}]$)} \\
Computation capacity $f_l$ [FLOPs/s] & 
\makecell[l]{Norm($\mu = 3\mathrm{e}{6}$, $\sigma = 1.1\mathrm{e}{7}$,\\ \quad $[2\mathrm{e}{6}, 6.5\mathrm{e}{6}]$)} \\
Computation capacity of GS/HAP $f_l$ [FLOPs/s] & 1.2e10,\ 0.5e10 \\
GU/AU transmit power $P$ [W] & 0.2,\ 0.15 \\
Active power of GS/HAP $P$ [W] & 120,\ 30 \\
Path loss factors $\eta_n^{\mathrm{LoS}},\ \eta_n^{\mathrm{NLoS}}$ & 0.1,\ 21 \\
LoS probability factors $\kappa_n^1,\ \kappa_n^2$ & 10,\ 0.6 \\
HAP altitude $h_{\mathrm{H}}$ [Km] & 20 \\
AU altitude $h_{\mathrm{u}}$ [m] & 200 \\
Carrier frequency $f_c$ [GHz] & 0.1 \\
Chip energy factor $\kappa$ & 1.2e10 \\
Max tolerable delay $\tau$ [s] & 2 \\
Delay cost weight $w$ & 0.3–0.7 \\
Noise power spectral density $\sigma^2$ [W/Hz] & 7.96159e–13 \\
System bandwidth $B$ [MHz] & 10 \\
\hline
\end{tabular}
}
\vspace{-5pt}
\end{table}

\subsubsection{Dataset Construction}
This study considers an LAE scenario covered by a single HAP. We generate eight training datasets of different scales using a heuristic-based Minimum Cost Maximum Flow (MCMF) graph algorithm. The MCMF algorithm formulates the task offloading problem as a flow representation and solves for the optimal task allocation and resource proportion. The configurations of the eight datasets are summarized in Table~\ref{tab:instance_scale}, where each dataset is named in the format “gsX\_guY\_auZ”, indicating that it contains $X$ GSs, $Y$ GUs, and $Z$ AUs.

We modified the heuristic MCMF algorithm to generate test data for our model by significantly increasing the number of heuristically generated weight combinations and applying brute-force search over the solution space to obtain optimal results. The algorithm is highly time-consuming, for example, generating a solution for a gs9\_gu19\_au12 instance takes about one minute. For each of the eight problem scales, we generated 2,000 test samples. All datasets were constructed on an Intel i7-13700F processor.

\begin{table}[t]
\small
\centering
\caption{Instance scale and generated sample count}
\label{tab:instance_scale}
\renewcommand{\arraystretch}{1.1}
\begin{tabular}{|>{\centering\arraybackslash}p{5.0cm}|>{\centering\arraybackslash}p{2.8cm}|}
\hline
\textbf{Scenario Scale} & \textbf{Generated Quantity} \\
\hline
$\rm gs2\_gu4\_au2$ & 70,000 \\
$\rm gs2\_gu5\_au3$ & 70,000 \\
$\rm gs3\_gu6\_au4$ & 80,000 \\
$\rm gs3\_gu7\_au5$ & 80,000 \\
$\rm gs6\_gu14\_au10$ & 80,000 \\
$\rm gs6\_gu16\_au11$ & 80,000 \\
$\rm gs9\_gu19\_au12$ & 90,000 \\
$\rm gs9\_gu20\_au13$ & 90,000 \\
\hline
\end{tabular}
\vspace{-15pt}
\end{table}
\begin{table}[t]
\small
\centering
\caption{Main hyperparameters of the model}
\label{tab:model_hyperparams}
\renewcommand{\arraystretch}{1.1}
\begin{tabular}{|>{\centering\arraybackslash}p{4.0cm}|>{\centering\arraybackslash}p{2.5cm}|}
\hline
\textbf{Parameter} & \textbf{Value} \\
\hline
learning\_rate & $1 \times 10^{-4}$ \\
diffusion\_type & gaussian \\
diffusion\_schedule & linear \\
diffusion\_steps & 200 \\
inference\_diffusion\_steps & 5 \\
inference\_trick & ddim \\
n\_layers & 10 \\
hidden\_dim & 256 \\
num\_attention\_heads & 4 \\
\hline
\end{tabular}
\end{table}

\subsection{Experimental Results}

\subsubsection{Model Configuration and Baseline Methods}
Table~\ref{tab:model_hyperparams} lists the main hyperparameters used in the GADSG algorithm. In addition, we compare GADSG against the following baseline methods:

\begin{itemize}
    \item \textbf{Random Execution (RE)}: Under constraint satisfaction, each task is randomly assigned to either local execution or offloading. For offloaded tasks, the target server allocates a random proportion of computing resources. Eight candidate solutions are randomly generated, and the one with the lowest cost is selected.

\item \textbf{Alternating Optimization (AO)}: A numerical method that optimizes the offloading decision and resource allocation in two separate stages. It first heuristically generates and fixes the resource allocation, then solves the offloading decision using a greedy strategy. Subsequently, with the offloading decision fixed, a convex optimization problem for resource allocation is formulated and solved using sequential quadratic programming (SQP).

\item \textbf{GRLO}~\cite{9592681}: A task offloading algorithm that combines Graph Neural Networks (GNNs) with reinforcement learning. It models the MEC system as a graph and utilizes GNNs to capture communication and computation relationships among nodes. Offloading strategies are learned through an Actor-Critic framework. In each decision round, GRLO generates offloading actions via graph embeddings and efficiently searches for near-optimal solutions using an enhanced sequential exploration algorithm.

\item \textbf{GDSG}~\cite{11006143}: A graph diffusion-based optimization algorithm that employs a Graph Convolutional Network (GCN) as its encoder. The GCN aggregates neighborhood features through weighted summation, mean, or max pooling, without relying on attention mechanisms. Control is solely achieved via contextual information encoded in the edge features.
\end{itemize}

Since our objective is to minimize the system's weighted cost, we adopt the following two evaluation metrics:

\begin{itemize}
    \item  \textbf{Average Cost Ratio}:  
Defined as the average ratio between the predicted cost produced by the model and the optimal cost across all test samples. On the test set, a value closer to 1 is preferred. Note that the ratio may occasionally fall below 1 because the test set does not exhaustively cover the solution space, and the model may occasionally produce solutions that outperform the reference.

\item \textbf{Cost Accuracy Rate}:  
Defined as the proportion of test samples where the model’s cost ratio is less than 1.1. This value is bounded above by 1, and higher values are better, indicating that the model can consistently produce high-quality offloading solutions across most task scenarios.
\end{itemize}

\subsubsection{Convergence and Inference Performance Comparison}
\begin{figure*}[htbp]
\centering
\subfigure[]{\includegraphics[width=0.247\textwidth]{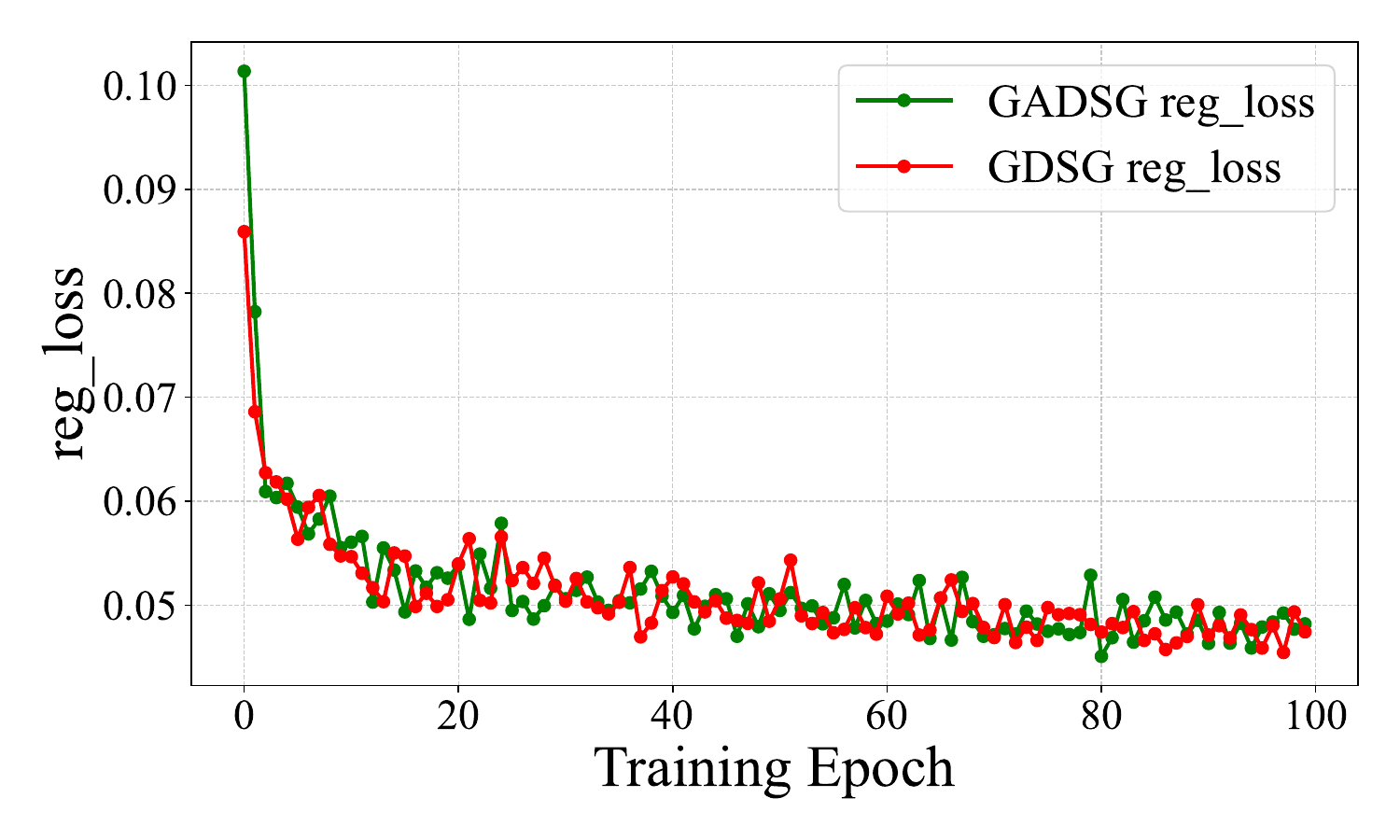}}\hfill
\subfigure[]{\includegraphics[width=0.247\textwidth]{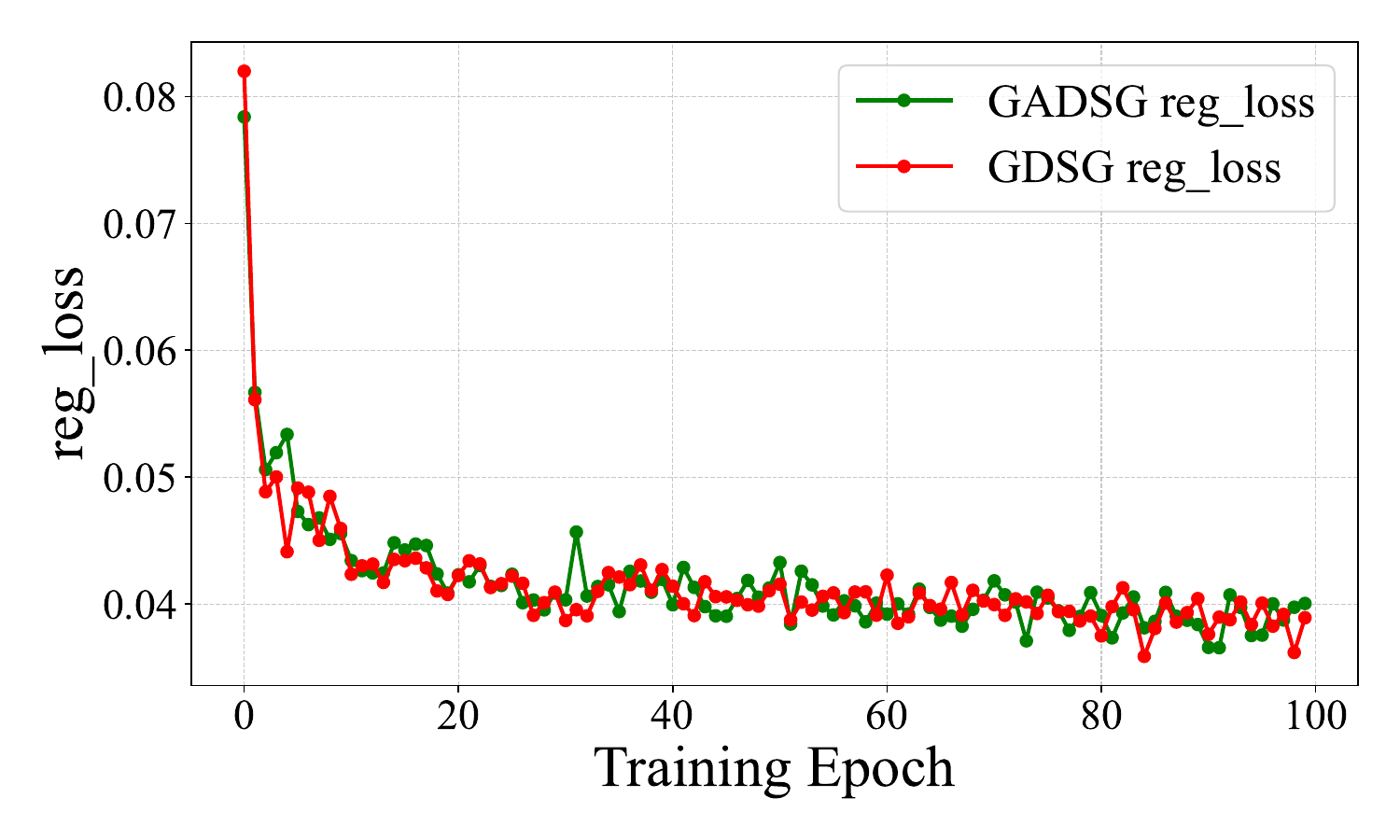}}\hfill
\subfigure[]{\includegraphics[width=0.247\textwidth]{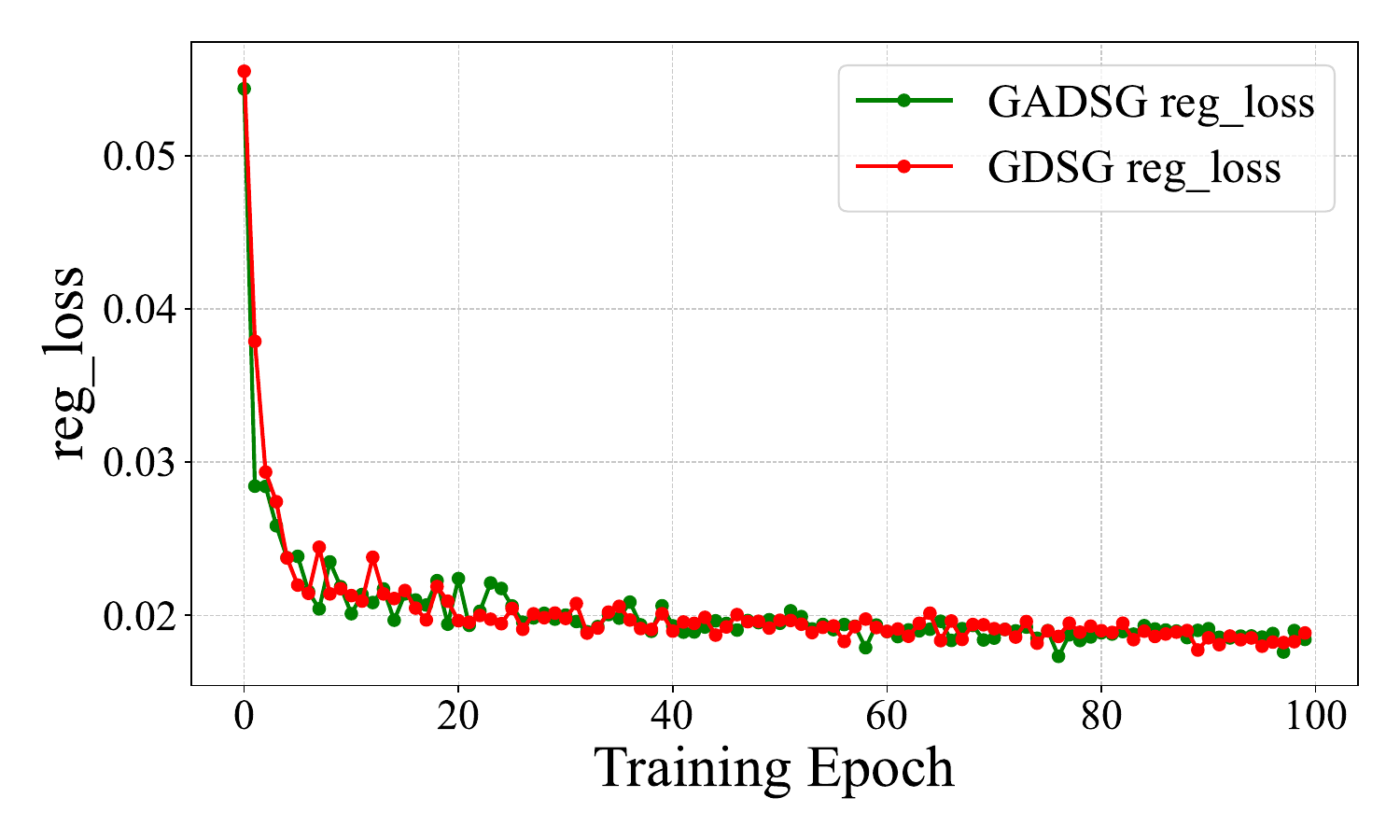}}\hfill
\subfigure[]{\includegraphics[width=0.247\textwidth]{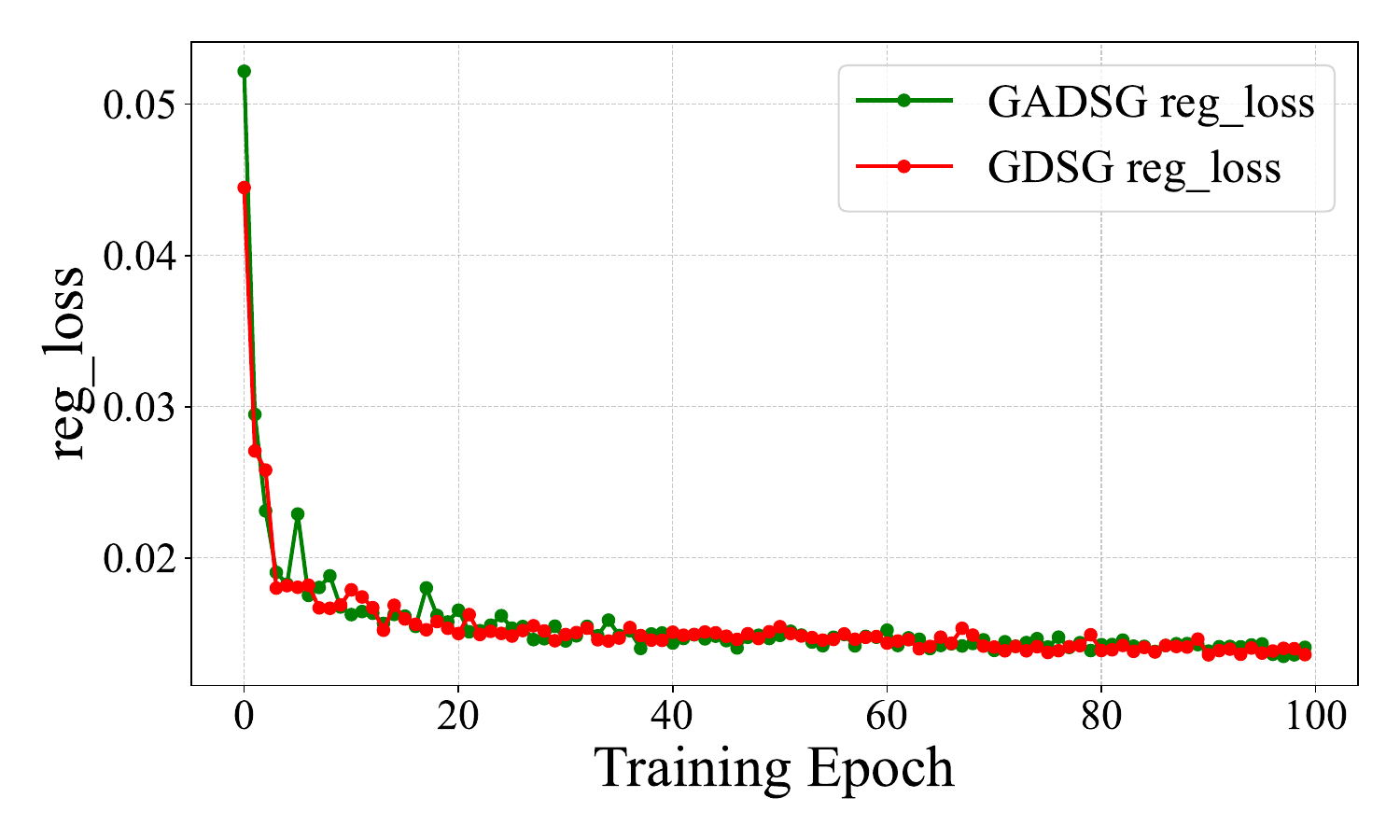}}
\caption{Convergence comparison between GADSG and GDSG during training, (a) on the$\rm gs2\_gu5\_au3$ dataset, (b) on the $\rm gs3\_gu6\_au4$ dataset, (c) on the $\rm gs6\_gu14\_au10$ dataset, and (d) on the $\rm gs9\_gu19\_au12$ dataset.}
\label{fig:loss}
\vspace{-10pt} 
\end{figure*}

This section evaluates the convergence behavior and inference performance of the GADSG model in comparison with the GDSG model. Experiments were conducted on four representative problem scales, and consistent trends were observed across all settings. As shown in Fig.~\ref{fig:loss}, the regression loss curves of GADSG and GDSG during training on the four datasets$\rm gs2\_gu5\_au3$, $\rm gs3\_gu6\_au4$, $\rm gs6\_gu14\_au10$, and $\rm gs9\_gu19\_au12$—are presented (classification loss curves are omitted due to their similar trends). It can be observed that both models exhibit rapid loss reduction in the early training stage and quickly converge to stable regions. Moreover, their overall loss values and convergence trajectories are nearly identical, indicating that incorporating GAT does not affect the convergence rate or training error control. The two models demonstrate similar fitting capabilities concerning the optimization objective, as introducing GAT does not alter the backpropagation optimization path during training.

\begin{figure*}[htbp]
\centering
\subfigure[]{\includegraphics[width=0.247\textwidth]{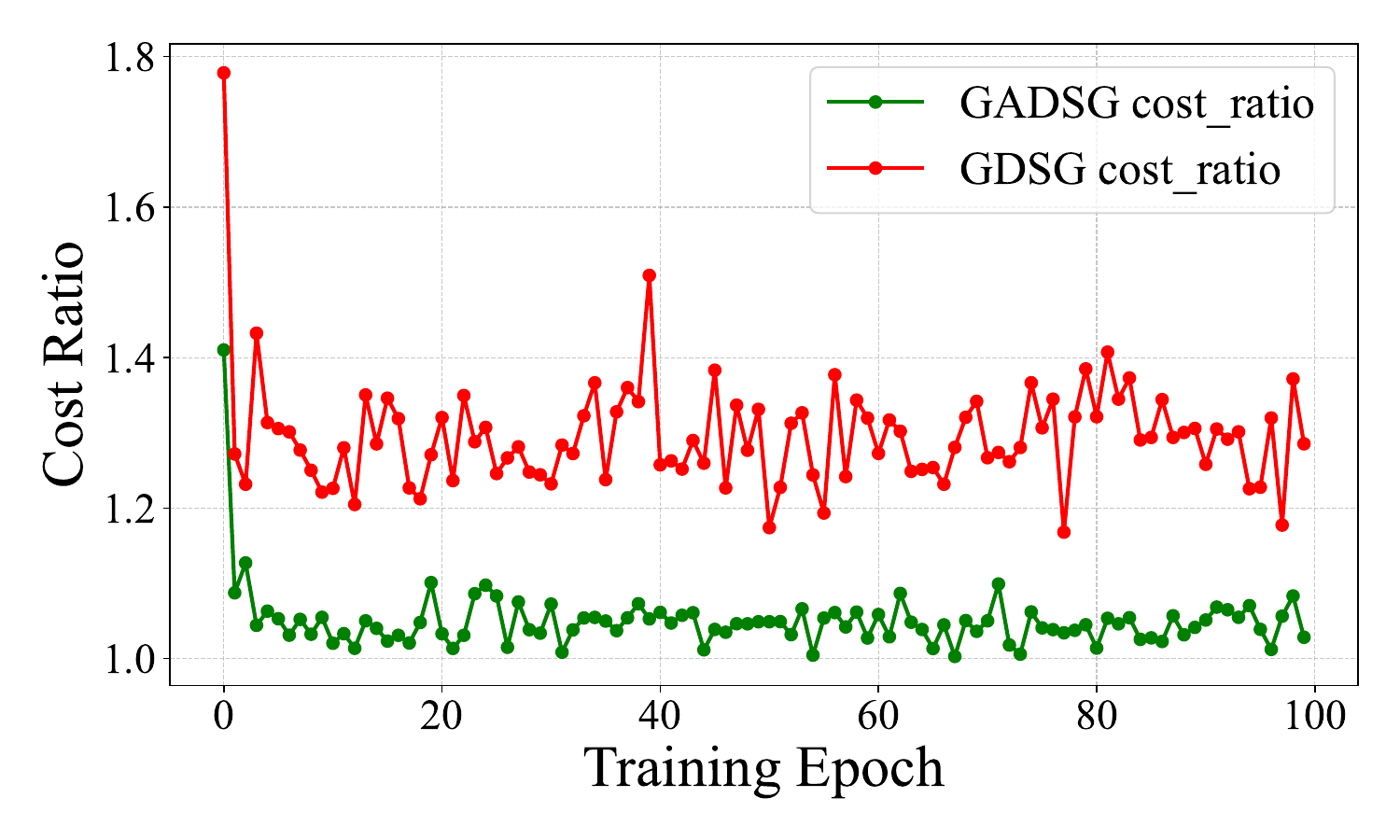}}\hfill
\subfigure[]{\includegraphics[width=0.247\textwidth]{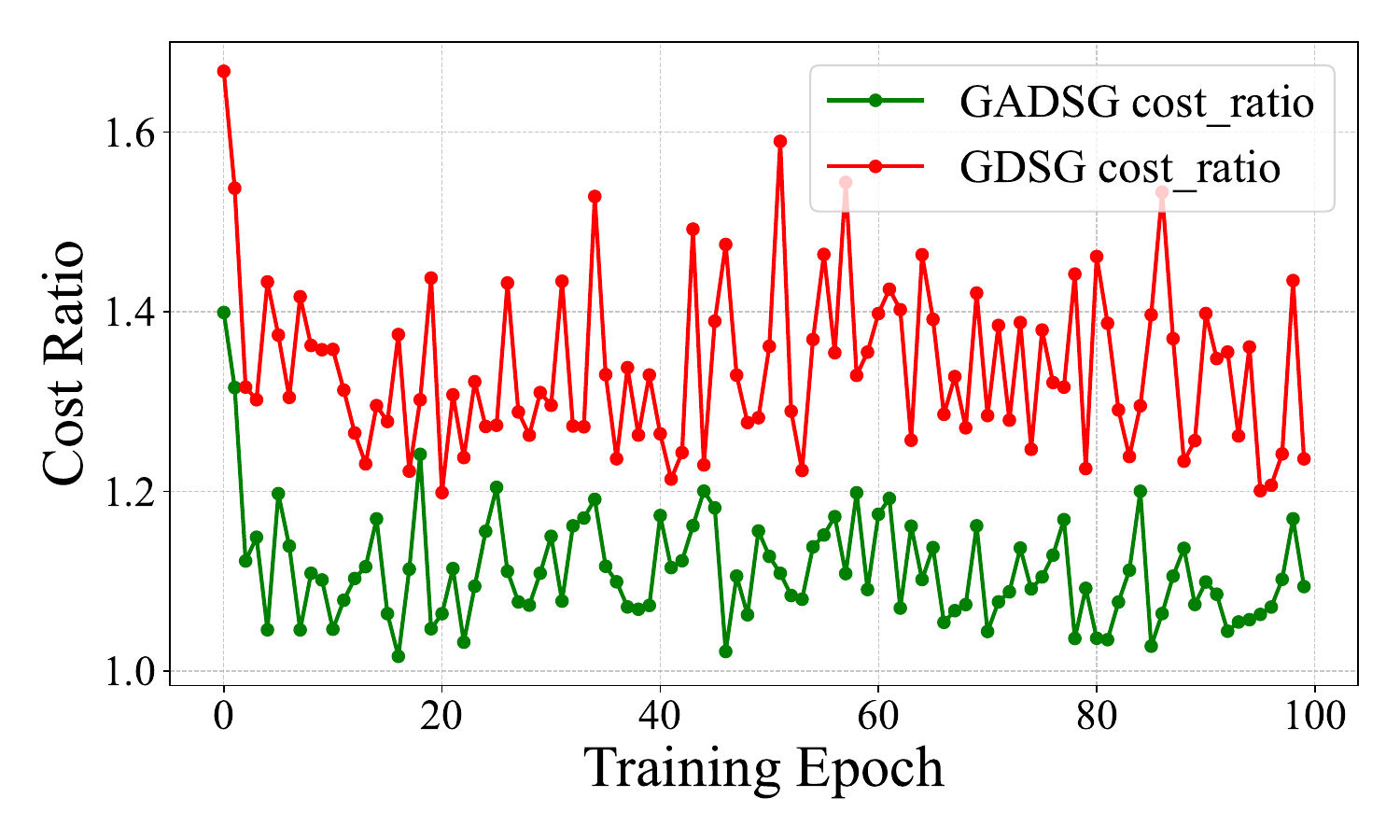}}\hfill
\subfigure[]{\includegraphics[width=0.247\textwidth]{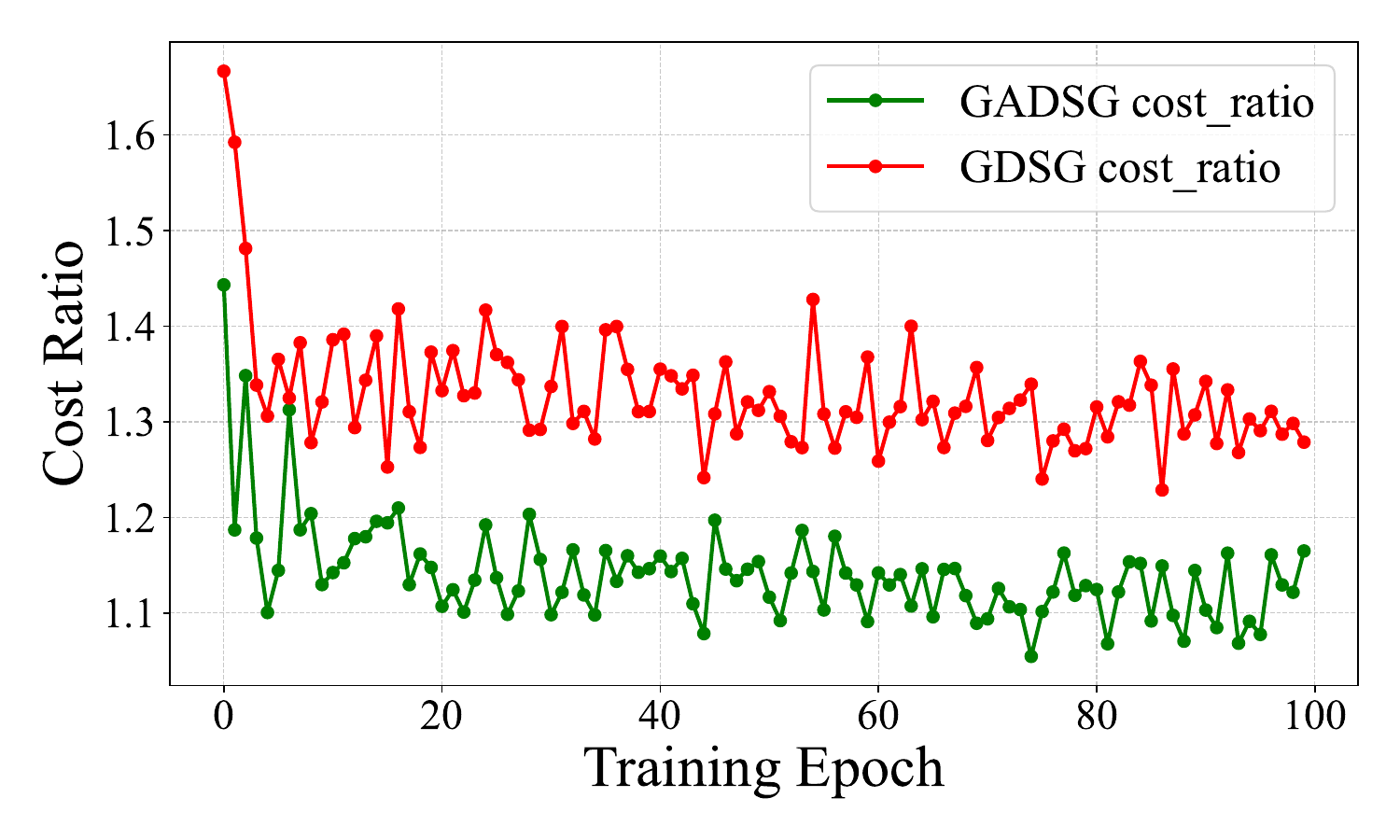}}\hfill
\subfigure[]{\includegraphics[width=0.247\textwidth]{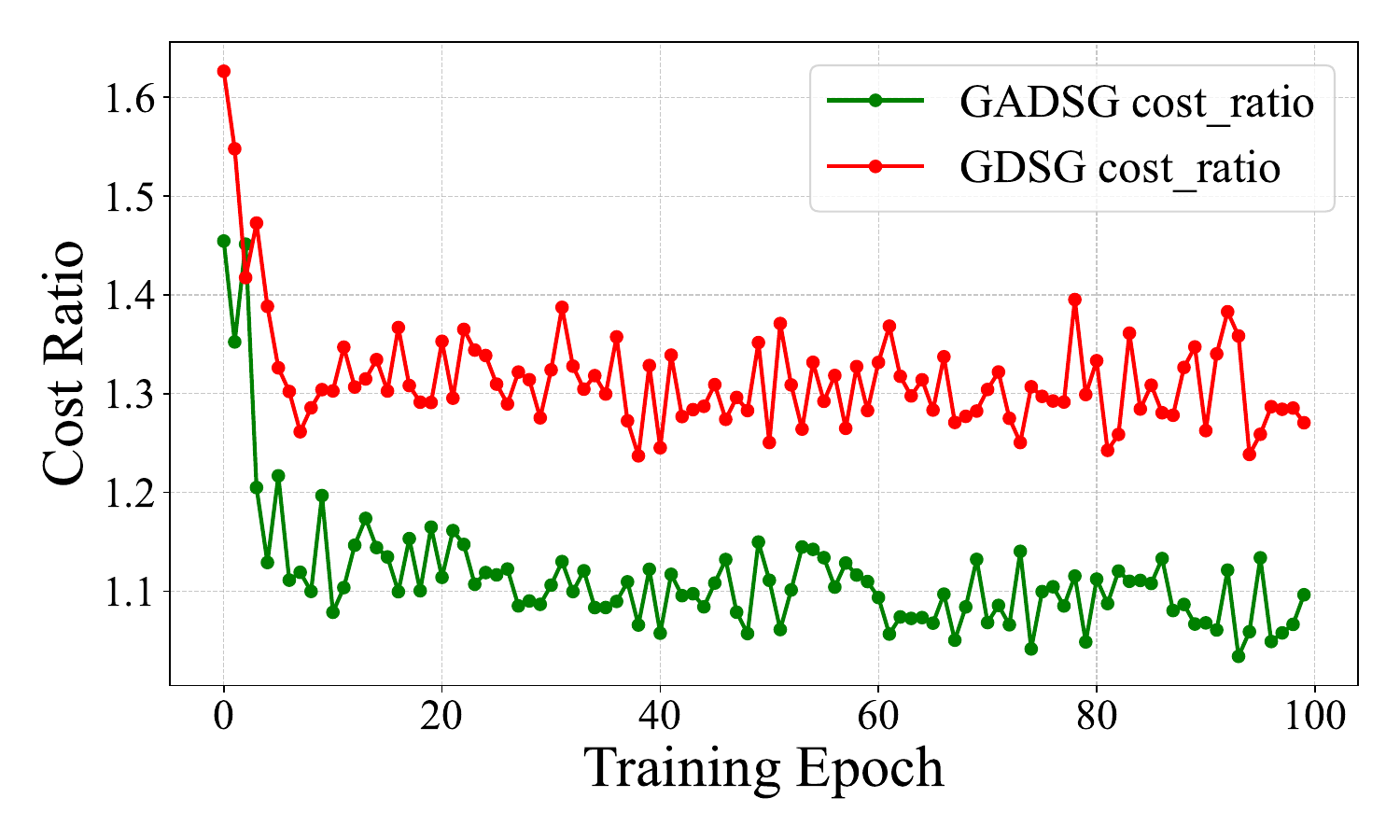}}
\caption{Cost ratio comparison between GADSG and GDSG during training, (a) on the $\rm gs2\_gu5\_au3$ dataset, (b) on the $\rm gs3\_gu6\_au4$ dataset, (c) on the $\rm gs6\_gu14\_au10$ dataset, and (d) on the $\rm gs9\_gu19\_au12$ dataset.}
\label{fig:cost}
\vspace{-10pt} 
\end{figure*}

However, a clear distinction emerges in the inference performance during model validation. As illustrated by the cost ratio curves in Fig.~\ref{fig:cost}, the GADSG model consistently maintains a lower cost ratio range throughout the training process, with relatively small fluctuations. In contrast, the GDSG model exhibits significantly higher predicted costs along with noticeable deviations and instability. This trend is consistently observed across all four datasets.

These results indicate that although the graph attention mechanism does not accelerate the training speed, its structural modeling advantages enable the model to produce task offloading solutions that are closer to optimal during inference. This further validates the critical role of attention mechanisms in selectively aggregating important edge information within complex heterogeneous graph structures.

\begin{table}[t]
\small
\centering
\caption{Comparison of average cost ratio↓ performance between GADSG and baseline methods}
\label{tab:cost_ratio}
\renewcommand{\arraystretch}{1.1}
\begin{tabularx}{\linewidth}{|>{\footnotesize}X|c|c|c|c|c|}
\hline
\textbf{Test Dataset} & \textbf{RE} & \textbf{AO} & \textbf{GRLO} & \textbf{GDSG} & \textbf{GADSG} \\
\hline
$\rm gs2\_gu4\_au2gt$  & 2.04 & 1.24 & 1.37 & 1.23 & \textbf{1.08} \\
$\rm gs2\_gu5\_au3gt$  & 2.12 & 1.27 & 1.42 & 1.21 & \textbf{1.06} \\
$\rm gs3\_gu3\_au4gt$  & 2.46 & 1.24 & 1.25 & 1.22 & \textbf{1.02} \\
$\rm gs3\_gu7\_au5gt$  & 2.35 & 1.23 & 1.29 & 1.15 & \textbf{1.05} \\
$\rm gs6\_gu14\_au10gt$ & 2.58 & 1.28 & 1.23 & 1.11 & \textbf{0.99} \\
$\rm gs6\_gu16\_au11gt$ & 2.79 & 1.31 & 1.25 & 1.12 & \textbf{0.97} \\
$\rm gs9\_gu19\_au12gt$ & 3.13 & 1.26 & 1.15 & 1.10 & \textbf{0.96} \\
$\rm gs9\_gu20\_au13gt$ & 3.02 & 1.25 & 1.17 & 1.11 & \textbf{0.98} \\
\hline
\end{tabularx}
\vspace{-15pt} 
\end{table}

\subsubsection{Performance and Robustness Comparison}

To comprehensively evaluate the optimization performance of the proposed GADSG algorithm in task offloading and resource allocation within LAENet-based MEC systems, we conduct comparative experiments against four baseline methods (RE, AO, GRLO, and GDSG) across eight test datasets of varying scales and topologies. Two evaluation metrics—average cost ratio and cost accuracy rate—are adopted to assess the optimization effectiveness and robustness of the algorithms. Table~\ref{tab:cost_ratio} presents a comparison of the average cost ratio achieved by different methods, while Table~\ref{tab:cost_accuracy} reports the results of the cost accuracy rate for each method.

\begin{table}[t]
\small
\centering
\caption{Comparison of cost accuracy rate↑ performance between GADSG and baseline methods}
\label{tab:cost_accuracy}
\renewcommand{\arraystretch}{1.1}
\begin{tabularx}{\linewidth}{|>{\footnotesize}X|c|c|c|c|c|}
\hline
\textbf{Test Dataset} & \textbf{RE} & \textbf{AO} & \textbf{GRLO} & \textbf{GDSG} & \textbf{GADSG} \\
\hline
$\rm gs2\_gu4\_au2gt$  & 0.13 & 0.75 & 0.64 & 0.77 & \textbf{0.91} \\
$\rm gs2\_gu5\_au3gt$  & 0.11 & 0.72 & 0.67 & 0.79 & \textbf{0.89} \\
$\rm gs3\_gu3\_au4gt$  & 0.14 & 0.76 & 0.62 & 0.83 & \textbf{0.96} \\
$\rm gs3\_gu7\_au5gt$  & 0.09 & 0.77 & 0.69 & 0.85 & \textbf{0.94} \\
$\rm gs6\_gu14\_au10gt$ & 0.06 & 0.69 & 0.80 & 0.90 & \textbf{1.00} \\
$\rm gs6\_gu16\_au11gt$ & 0.09 & 0.66 & 0.77 & 0.82 & \textbf{1.00} \\
$\rm gs9\_gu19\_au12gt$ & 0.03 & 0.76 & 0.87 & 0.89 & \textbf{1.00} \\
$\rm gs9\_gu20\_au13gt$ & 0.05 & 0.71 & 0.86 & 0.88 & \textbf{1.00} \\
\hline
\end{tabularx}
\vspace{-15pt} 
\end{table}

Specifically, the RE algorithm performs the worst across all task scenarios, with an average cost ratio significantly higher than other methods and a very low cost accuracy rate. In LAENet-based MEC environments characterized by high heterogeneity and unbalanced link quality, random offloading strategies often lead to resource mismatch or server overload, resulting in extremely poor solution quality.

The AO algorithm shows moderate performance on both metrics. Due to its reliance on fixed initial strategies during optimization, it tends to fall into local optima. Furthermore, AO struggles to effectively coordinate global strategies for offloading and resource allocation in highly coupled multi-task scenarios, thereby limiting its overall performance.

The GRLO algorithm performs slightly better than AO in some test scenarios, indicating that graph-based reinforcement learning possesses a certain degree of structural awareness and policy learning capability. However, its average cost ratio exhibits significant fluctuations, and the cost accuracy rate varies considerably across datasets. This instability arises from two primary factors. First, reinforcement learning is highly sensitive to the state space and training distribution, resulting in poor policy stability in complex network structures or scenarios with dynamic task topologies. Second, GRLO generates offloading strategies based primarily on local graph embeddings during the policy output phase. As a result, its optimization process focuses more on local action exploration. In contrast to graph diffusion-based models, GRLO struggles to perform truly global optimization across the graph structure.

\begin{figure*}[t]
\centering
\subfigure[]{\includegraphics[width=0.480\textwidth]{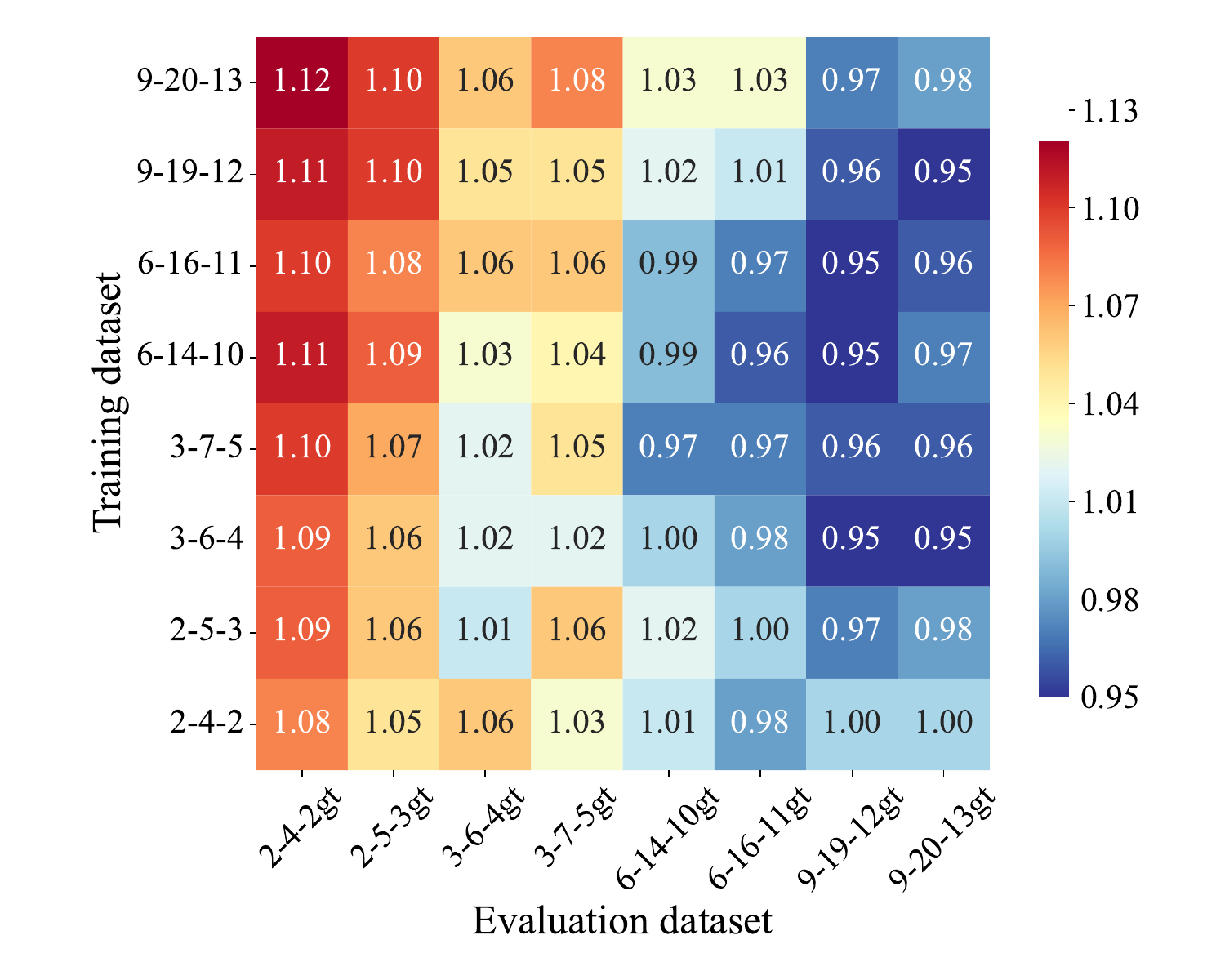}}\hfill
\subfigure[]{\includegraphics[width=0.480\textwidth]{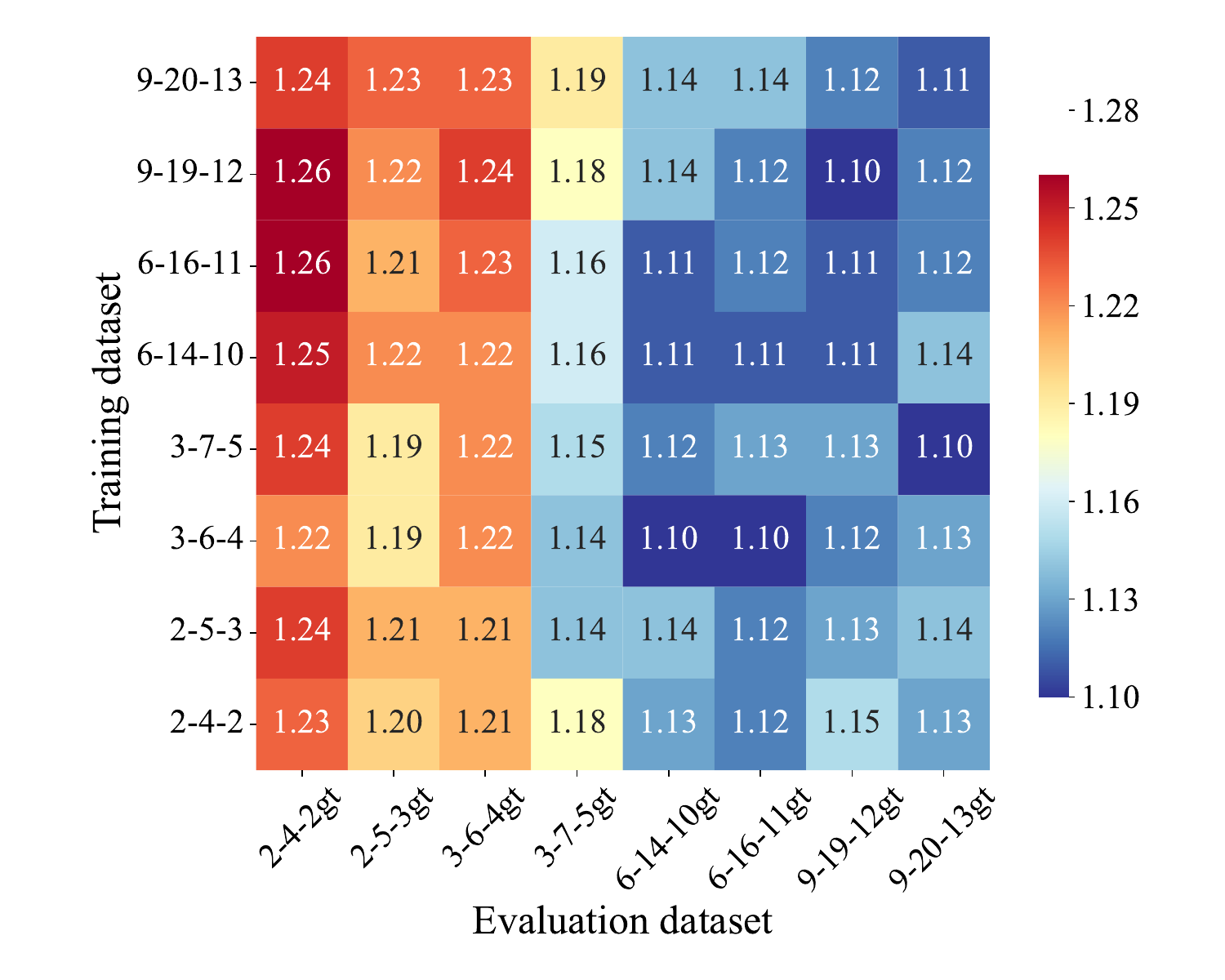}}
\caption{Comparison of average cost ratio across problem scales, where the proposed GADSG is shown in (a) and GDSG is shown in (b).}
\label{fig:two_merged}
\vspace{-15pt} 
\end{figure*}

As a graph diffusion method with more thorough graph structural modeling, the GDSG algorithm demonstrates overall stable performance, with the average cost ratio and cost accuracy rate significantly outperforming AO and GRLO. However, it still falls short compared to the proposed GADSG algorithm. Furthermore, as observed in the original GDSG literature~\cite{11006143}, the problem setting was based on a multi-server multi-user MEC system with homogeneous node types and relatively regular link topology. Under similar scenario scales, the average cost ratio reported in the original experiments is lower than that in our experiments on LAE-based MEC systems. This discrepancy is attributed to the increased complexity of task graph structures in LAE networks. Since GDSG relies on fixed aggregation functions such as averaging or max-pooling to fuse neighborhood information, it cannot dynamically model the importance of nodes and links, making it less capable of producing optimal offloading decisions.

The proposed GADSG method consistently achieved the best performance across all test datasets, exhibiting the lowest average cost ratio and reaching a 100\% cost accuracy rate in complex scenarios, which demonstrates its exceptional optimization capability and robustness. Notably, on several large-scale datasets, the cost ratio achieved by GADSG was even less than 1, since the trained GADSG model can approximate the optimal solution more closely than the test labels through a limited number of parallel samplings. By integrating the graph diffusion mechanism with the GAT, GADSG assigns learnable weights to adjacent nodes and links during the diffusion process, enabling dynamic modeling of critical connections in heterogeneous topologies. Compared with the static aggregation strategy adopted in GDSG, GADSG significantly enhances structural representation and decision-generation flexibility, allowing it to produce near-optimal solutions even under increasing structural complexity consistently.

\subsubsection{Generalization Performance Comparison}
Considering the highly dynamic nature of low-altitude economy (LAE) networks, where task structures, user scales, and resource configurations frequently fluctuate over time and across scenarios, the model must possess strong cross-scale transferability. Therefore, we further design a cross-scale evaluation experiment to investigate the generalization performance of both GADSG and GDSG methods.

The cross-scale performance comparison of average cost ratio for the two methods is shown in Fig.~\ref{fig:two_merged}. Overall, GADSG significantly outperforms GDSG in generalization across problem sizes. It consistently maintains the average cost ratio between 0.95 and 1.13 on most test sets, showing more balanced results. GDSG relies on static aggregation in graph convolution, which limits its ability to adapt to topological changes, thus reducing its generalization capacity. In contrast, GADSG integrates multi-head attention into the graph diffusion process, enabling dynamic modeling of node importance and global structure. This allows the model to generate high-quality offloading and resource allocation strategies even on task graphs with structures differing significantly from the training set.

\section{Conclusion} \label{conclusion}
This paper investigates the optimization of task offloading and resource allocation in mobile edge computing (MEC) systems under low-altitude economy (LAE) networks with multi-type nodes and dynamic communication links. First, we construct a three-layer heterogeneous LAENet-based MEC architecture with air-ground coordination. The system is modeled from the perspectives of communication channels, computational cost, and constraint conditions, and the problem is uniformly abstracted as a joint optimization task over a graph structure. To enhance the modeling capacity for complex topologies and improve optimization performance, we propose a graph-attention-based diffusion sampling algorithm named GADSG. This method models and jointly optimizes the discrete offloading decisions and continuous resource allocation ratios in the latent space. By combining the structural generalization ability of diffusion models with the context-aware attention mechanism, GADSG is capable of generating constraint-satisfying and high-quality solutions.

To validate the performance of the proposed algorithm, we construct eight LAE-based MEC task graph datasets of varying scales and conduct comparative experiments against multiple baseline methods in terms of convergence, optimization accuracy, robustness, and generalization. The results show that GADSG consistently achieves superior performance across all task scenarios. In particular, it demonstrates stronger adaptability in cross-scale transfer tests, further confirming its potential for deployment in dynamic and complex LAE network environments. 
\vspace{-2mm}




\bibliographystyle{IEEEtran}  
\bibliography{references}


 




\vfill

\end{document}